\documentclass{jfm}
\usepackage{graphicx,ulem}
\usepackage{epstopdf, epsfig}
\usepackage{amsmath}
\usepackage{amsfonts}
\usepackage{amssymb}
\usepackage{multirow}

\usepackage{xcolor}

\newcommand{\diff}{\text{d}}

\usepackage{psfrag}
\usepackage{tikz}
\usetikzlibrary{arrows}
\usetikzlibrary{plotmarks}
\usetikzlibrary{external}
\usetikzlibrary{patterns}
\tikzset{external/force remake=true}

\usepackage{pgfplots}
\usepgfplotslibrary{groupplots}
\usepgfplotslibrary{fillbetween}
\pgfplotsset{compat=newest}
     \newlength\fheight 
     \newlength\fwidth
     \newlength\svgwidth

\shorttitle{Triadic resonant instability in confined and unconfined axisymmetric geometries}
\shortauthor{S. Boury, P. Maurer, S. Joubaud, T. Peacock, P. Odier}

\title{Triadic resonant instability in confined and unconfined axisymmetric geometries}

\author{S. Boury\aff{1,2,3}
  \corresp{\email{sb7918@nyu.edu}}, 
  P. Maurer\aff{2},
  S. Joubaud\aff{2,4},
  T. Peacock\aff{3},
  P. Odier\aff{2}}
  
\affiliation{\aff{1}Courant Institute of Mathematical Sciences, New York University, New York, NY 10012, USA
\aff{2}ENS de Lyon, CNRS, Laboratoire de Physique, F-69342 Lyon, France
\aff{3}Department of Mechanical Engineering, Massachusetts Institute of Technology, Cambridge, MA 02139, USA
\aff{4}Institut Universitaire de France (IUF), 1 rue Descartes 75005 Paris, France}

\begin{document}

\maketitle

\begin{abstract}

	We present an investigation of the resonance conditions of axisymmetric internal wave sub-harmonics in confined and unconfined domains. In both cases, sub-harmonics can be spontaneously generated from a primary wave field if they satisfy at least a resonance condition on their frequencies, of the form $\omega_0 = \pm \omega_1 \pm \omega_2$. We demonstrate that, in an unconfined domain, the sub-harmonics follow three dimensional spatial resonance conditions similar to the ones of Triadic Resonance Instability (TRI) for Cartesian plane waves. In a confined domain, however, the spatial structure of the sub-harmonics is fully determined by the boundary conditions and we observed that these conditions prevail upon the resonance conditions. In both configurations, these findings are supported by experimental data showing good agreement with analytical and numerical derivations.
\end{abstract}

\begin{keywords}
Authors should not enter keywords on the manuscript, as these must be chosen by the author during the online submission process and will then be added during the typesetting process (see http://journals.cambridge.org/data/\linebreak[3]relatedlink/jfm-\linebreak[3]keywords.pdf for the full list)
\end{keywords}

%
%
%
%
%

	\section{Introduction}

		Ubiquituous in the oceans and in the atmosphere, internal wave studies share a long and rich history, since the preliminary works of \cite{gortler1943} and \cite{mowbray1967b}. Non-linear interactions of internal waves are relevant to the ocean understanding as they participate to energy transfer between scales. These processes have been studied in the case of several wave fields interacting together \citep[e.g.][]{husseini2019} and in the case of self-interacting wave fields \citep[e.g.][]{baker2020}. Self-interaction can be categorised into two separate dual mechanisms \citep{bouryPhD}: Super-Harmonic Generation (SHG) \citep[e.g.][]{baker2020, varma2020, boury2020a}; and generation of sub-harmonics via Triadic Resonant Instability (TRI) \citep[e.g.][]{tabaei2003, joubaud2012, bourget2013, karimi2014, kataoka2015, richet2017, sarkar2017, dauxois2018}. The latter case has been observed in various configurations and has been widely studied in two-dimensional Cartesian geometry~\citep{benielli1998, mcewan1971, mcewan1972, staquet2002, thorpe1968a, joubaud2012}. TRI is characterised by the non-linear breaking of an energetic internal wave (usually called \textit{primary wave}) into two waves of smaller frequency (the sub-harmonic \textit{secondary waves}) \citep{joubaud2012, bourget2013, maurer2016}. The resulting triad of monochromatic plane waves formed is in resonance, which means that their frequencies $\omega$ and wave vectors $\mathbf{k}$ are linked through linear relations
		\begin{eqnarray}
			\omega_0 &=& \pm \omega_1 \pm \omega_2,
			\label{intro:eq1}\\
			\mathbf{k_0} &=& \pm \mathbf{k_1} \pm \mathbf{k_2},
			\label{intro:eq2}
		\end{eqnarray}
		where $\omega_0$ and $\mathbf{k_0}$ (respectively $\omega_1$, $\omega_2$ and $\mathbf{k_1}$, $\mathbf{k_2}$) are related to the primary wave (respectively the secondary waves). Note that, in these experiments, when the excitation is a mode (i.e. a horizontal standing wave propagating vertically, or the contrary), the two sub-harmonics do not necessary have a modal structure \citep{joubaud2012}.
	
		Most of the studies on these non-linear processes have been conducted in two-dimensional Cartesian geometry, based on plane wave formalism. Although recent works have explored the non-trivial implications of a three-dimensional domain (see e.g. \cite{mora2021}), the cylindrical counterpart is poorly documented. Experiments conducted in axisymmetric geometry have shown, however, that changing geometry allows for a rich dynamics and a wide variety of interesting non-linear behaviours \citep{maurerPhD, shmakova2018, boury2020a, boury2020b}. For example, \cite{boury2020a} have shown that super-harmonics can be spontaneously generated in axisymmetric geometry via self-interaction of a monochromatic wave field, in a non-rotating linearly stably stratified fluid. Through the investigation of an axisymmetric inertial wave attractor, \cite{sibgatullin2017} and \cite{boury2020b} have also shown that symmetry breakings are likely to occur with high energetic wave fields, exhibiting loss of axisymmetry and forming periodic patterns in the equatorial plane, often disregarded.
		
		Spontaneous generation of internal wave sub-harmonics in axisymmetric geometry has indeed been observed in a few experimental studies. The existence of resonant triads in cylindrical geometry has been explored in the case of the elliptic instability triggered by a precessional forcing~\citep{albrecht2015, albrecht2018, eloy2003, giesecke2015, lagrange2011, lagrange2016, meunier2008}. In the case of a conical wave field forced by a vertically oscillating torus in a density stratified flow, \cite{shmakova2018} reported an experimental evidence of localised TRI, triggered at the apex of the internal wave cone, i.e. the focusing region. With a similar setup of an annular wave generator in non-uniform stratification, \cite{maurerPhD} also observed the generation of sub-harmonics, accompanied by a symmetry breaking, at the convergence point of the three-dimensional wave field. In both cases, a resonance condition on the frequencies is found to be satisfied by the forcing wave field and the two sub-harmonics, as well as on the vertical wave number. Observations differ on the radial resonance condition, seen to be satisfied locally in the case of \cite{shmakova2018} but not in the experiments of \cite{maurerPhD}, in which the symmetry breaking with non-zero radial velocity at the center of the domain prevents a simple radial resonance condition from existing. To this date, no exact derivation of the resonance conditions in cylindrical geometry has been proposed. This is essentially due to the analytical expression of the wave fields in terms of Bessel functions, that does not straightforwardly lead to a similar triadic condition as in Cartesian geometry. The present study is built upon these observations, and delves further into the problematic of sub-harmonic generation of internal waves in cylindrical geometry.

		We present analytical investigations of the resonance conditions of internal wave triads spontaneaously generated from a primary excitation field in confined and unconfined geometries, and we support them with experimental observations. This paper is organised as follows. The experimental apparatus is presented in section~\ref{sec:methods}. In section~\ref{sec:theory}, we derive the general equations governing internal waves in cylindrical geometry and present their linear solutions. The two different configurations of unconfined geometry and doubly-confined domain are then explored theoretically, with experimental verification, in sections~\ref{sec:unconfined} and~\ref{sec:confined}, respectively. Section~\ref{sec:conclusions} presents our conclusions and discussion on the present problem.

	\section{Methods}
	\label{sec:methods}
	
		\subsection{Numerical Tools}
		
			Bessel functions, Bessel integrals, and Fresnel integrals were evaluated using Matlab's functions and integration scheme, providing good estimates of these different quantities. When studying the dependence of Bessel function integrals on radial wave numbers, a resolution of $10^{-2}\mathrm{~m^{-1}}$ was used.
	
		\subsection{Geometries}
			
			Two different geometries are investigated in our study, both axisymmetric: an unconfined and a confined geometry. In the unconfined geometry, waves are free to propagate in an unbounded domain, whereas in the confined geometry, the accessible domain is restricted to a regular cylinder whose generatrix is vertical. In the experiments, as described below, it is important to note that the fluid is of course always bounded: in the case in which they are located sufficently far from the source, however, the boundaries have no impact on the non-linear behaviour of the waves, leading to the configuration that we call ``unconfined'', by opposition to the modal case in which the waves are constantly reflecting on the cylindrical boundaries, that we call ``confined''.
	
		\subsection{Experimental Apparatus}
		
\begin{figure}
			\centering
			\includegraphics[scale=0.8]{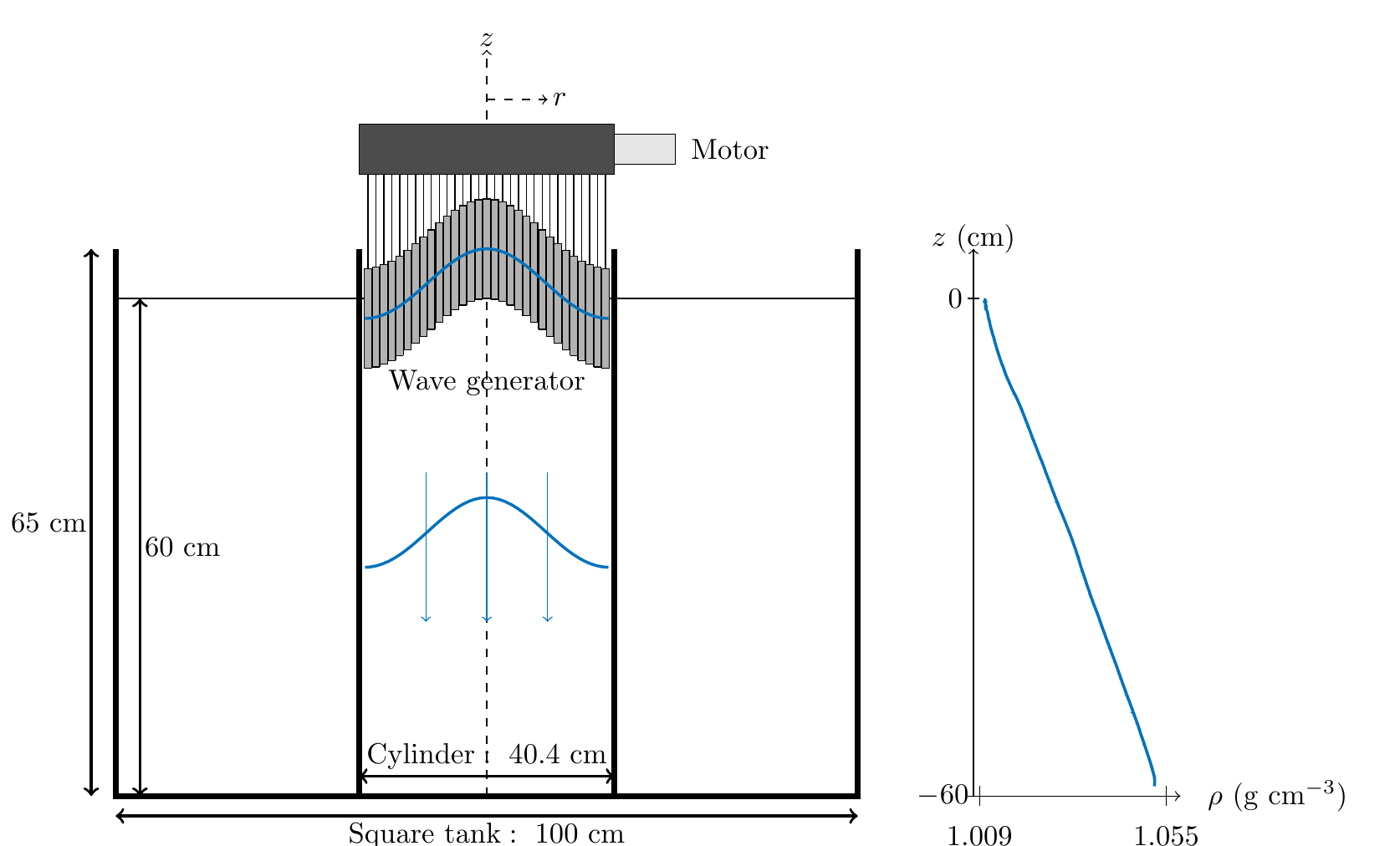}
			\caption{Left: schematic of the experimental apparatus in the confined configuration. The inner cylinder can be removed to perform experiments in a ``quasi'' unconfined geometry. Right: example of a linear stratification, from the confined experiments.}
			\label{fig:experiment}
		\end{figure}

			Complementary experiments were conducted using the experimental apparatus described in \cite{boury2018} and presented in figure~\ref{fig:experiment}. A $600~\mathrm{L}$ square base acrylic tank is filled with salt stratified water by using the double-bucket method, in order to obtain a linear stratification (constant vertical density gradient) \citep{fortuin1960, oster1963}.
			
			Internal wave fields were produced thanks to an axisymmetric wave generator \citep{maurer2017} located at the top of the tank. This device, adapted from two dimensional wave makers \citep{gostiaux2006}, is constituted of $16$ concentric cylinders oscillating vertically at the same tunable frequency and whose amplitudes can be set separately. Its reliability in producing diverse axisymmetric wave fields such as Bessel modes or conical beams has already been discussed in previous studies \citep{maurer2017, boury2018, boury2019a, boury2020a}.
			
			An additional acrylic cylinder of the same diameter as the generator is used to confine the wave field, hence two different sets of experiments were conducted: a first set without the cylindrical boundary; and a second set with it.
			
			Velocity fields were visualised thanks to a Particle Image Velocimetry (PIV) technique. Vertical and horizontal laser sheets were generated thanks to a $2\mathrm{~W}$ Ti:Sapphire laser (wavelength $532\mathrm{~nm}$) and a cylindrical lens. For the purpose of visualisation, the fluid was seeded with hollow glass spheres and silver coated spheres, both of $10\mathrm{~\mu m}$ diameter. Particle displacements were recorded at $4\mathrm{~Hz}$ using a camera located either on the side of the tank (visualisation in the vertical cross-section) or facing a $45^\circ$ mirror placed under the tank (visualisation in the horizontal cross-section). PIV raw images were processed thanks to the CIVx algorithm in order to extract the velocity fields \citep{fincham2000}.
	
	\section{Theory}
	\label{sec:theory}
	
		\subsection{Governing Equations}
		
			We consider an incompressible stratified fluid of density $\rho = \bar{\rho} + \rho'$, $\bar{\rho}$ being the background density and $\rho'$ its fluctuations, set in solid rotation at an angular velocity $\Omega$. In the Boussinesq approximation, the Navier-Stokes and conservation equations read in cylindrical coordinates
			\begin{eqnarray}
				\partial_t \mathbf{v} + (\mathbf{v} \cdot \mathbf{\nabla}) \mathbf{v} &=& - f \mathbf{e_z} \times \mathbf{v} - \frac{1}{\rho_0}\mathbf{\nabla}P + \mathbf{b},\label{eq:eq1}\\
				\partial_t \mathbf{b} + (\mathbf{v} \cdot \mathbf{\nabla}) \mathbf{b} &=& - N^2 v_z \mathbf{e_z},\label{eq:eq2}\\
				\mathbf{\nabla}\cdot\mathbf{v} &=& 0.\label{eq:eq3}
			\end{eqnarray}
			with $\mathbf{v} = (v_r, v_\theta, v_z)$, $\mathbf{b} = b\mathbf{e_z}$, and $P$ the velocity, buoyancy, and pressure fields, respectively. Here we define the buoyancy $b$ and the buoyancy frequency $N$ as
			\begin{equation}
				b = -g \frac{\rho'}{\rho_0} \mathrm{~~~~~~~and~~~~~~~} N(z) = \sqrt{-\frac{g}{\rho_0}\left.\frac{\partial \bar{\rho}}{\partial z}\right|_z},\label{eq:eq4}
			\end{equation}
			with $\rho_0$ the mean density, and the Coriolis frequency $f$ as $f=2\Omega$. We introduce the vorticity $\mathbf{\xi}$ as
			\begin{equation}
				\mathbf{\xi} = \mathbf{\nabla}\times\mathbf{v}.\label{eq:eq5}
			\end{equation}
			Hence, taking the curl of equation~\eqref{eq:eq1} eliminates the pressure term $\mathbf{\nabla} P$ and allows us to write, after some algebra,
			\begin{equation}
				\partial_t \mathbf{\xi} +  \mathbf{\nabla} \times \left[ (\mathbf{v} \cdot \mathbf{\nabla}) \mathbf{v}\right] = f \partial_z \mathbf{v} + \mathbf{\nabla}\times(b \mathbf{e_z}).\label{eq:eq6}
			\end{equation}
			Equations \eqref{eq:eq2} and \eqref{eq:eq6} collapse together using the time derivative of \eqref{eq:eq6} and the curl of \eqref{eq:eq2}, leading to
			\begin{equation}
				\partial^2_t \mathbf{\xi} + \mathcal{N} (\mathbf{v},\mathbf{b}) = f \partial_t\partial_z \mathbf{v} - N^2 \mathbf{\nabla}\times(v_z \mathbf{e_z}),\label{eq:eq7}
			\end{equation}
			with the non-linear terms
			\begin{equation}
				\mathcal{N} (\mathbf{v},\mathbf{b}) = \mathbf{\nabla} \times \left[ \partial_t \left( (\mathbf{v} \cdot \mathbf{\nabla}) \mathbf{v}\right) + (\mathbf{v} \cdot \mathbf{\nabla}) \mathbf{b} \right].\label{eq:eq8}
			\end{equation}			
			Then, the curl of \eqref{eq:eq7} allows us to write the non-linear equation for vorticity
			\begin{equation}
				\partial^2_t \mathbf{\nabla}\times\mathbf{\xi} + \mathbf{\nabla}\times\mathcal{N} (\mathbf{v},\mathbf{b}) = f \partial_t\partial_z \mathbf{\xi} - N^2 \mathbf{\nabla}\times\left[\mathbf{\nabla}\times(v_z \mathbf{e_z})\right].\label{eq:eq9}
			\end{equation}

		\subsection{Solutions of the Linear Problem}
				
			We first consider low amplitude waves, for which non-linear effects can be neglected. Hence, the linearised equations are obtained by setting the non-linear term $\mathcal{N}$ in equation~\eqref{eq:eq9} to zero, leading to the following time evolution equation
			\begin{equation}
				\partial^2_t \mathbf{\nabla}\times\mathbf{\xi} = f \partial_t\partial_z \mathbf{\xi} - N^2\mathbf{\nabla}\times\left[ \mathbf{\nabla}\times(v_z \mathbf{e_z})\right].\label{eq:eq10}
			\end{equation}			
			Using the volume conservation equation~\eqref{eq:eq3}, the curl of the vorticity simply writes in terms of the Laplacian of the velocity. Equation \eqref{eq:eq10} writes
			\begin{equation}
				\partial^2_t \Delta^\odot \mathbf{v} = - f\partial_t\partial_z \mathbf{\xi} + N^2 \mathbf{\nabla}\times\left[\mathbf{\nabla}\times(v_z \mathbf{e_z})\right],\label{eq:eq11}
			\end{equation}
			in terms of the velocities $v_r$, $v_\theta$, and $v_z$. The vectorial cylindrical Laplacian $\Delta^\odot$ is defined involving a coupling between the radial and orthoradial velocities $v_r$ and $v_\theta$ as
			\begin{equation}
				\Delta^\odot \mathbf{v} = 
				\begin{bmatrix}
					\Delta v_r -\frac{1}{r^2} \left( v_r + 2 \partial_\theta v_\theta \right) \\
					\Delta v_\theta -\frac{1}{r^2} \left( v_\theta - 2 \partial_\theta v_r \right)\\
					\Delta v_z
				\end{bmatrix},\label{eq:eq12}
			\end{equation}
			with the scalar cylindrical Laplacian
			\begin{equation}
				\Delta v = \frac{1}{r}\partial_r \left( r \partial_r v \right) + \frac{1}{r^2} \partial^2_\theta v + \partial^2 _z v.\label{eq:eq13}
			\end{equation}
			Therefore, the linearisation of equation~\eqref{eq:eq9} yields
			\begin{eqnarray}
				\partial_t^2 \Delta v_z + f^2 \partial^2_z v_z + N^2 \left( \frac{1}{r}\partial_r (r \partial_r v_z)  + \frac{1}{r^2}\partial^2_\theta v_z \right) &=& 0,\label{eq:eq14}\\
				\partial_t^2 \Delta v_r -\frac{1}{r^2}\partial_t^2\left( v_r + 2\partial_\theta v_\theta \right) + f \partial_t \partial_z \left( \frac{1}{r}\partial_\theta v_z - \partial_z v_\theta \right) - N^2 \partial_r \partial_z v_z &=& 0,\label{eq:eq15}\\
				\partial_t^2 \Delta v_\theta -\frac{1}{r^2}\partial_t^2\left( v_\theta - 2\partial_\theta v_r \right) + f \partial_t \partial_z \left( \partial_z v_r - \partial_r v_z \right) - N^2 \frac{1}{r}\partial_\theta \partial_z v_z &=& 0.\label{eq:eq16}
			\end{eqnarray}
			
			This system can be solved analytically, and its solutions are described by the following functions, called Kelvin modes		
				\begin{eqnarray}
					v_r (r, \theta , z, t) &=& i \frac{m v_z^0}{4 l \omega}\left[ (f-2\omega) J_{p-1} (l r) + (f + 2\omega) J_{p+1} (l r) \right] e^{i (\omega t - m z - p \theta)} + \mathrm{c.c.},~~~~~\label{eq:eq17}\\
					v_\theta (r, \theta , z, t) &=& \frac{m v_z^0}{2 l \omega} \left[ (2 f-\omega) J_{p-1} (l r) - (2 f + \omega) J_{p+1} (l r) \right] e^{i (\omega t - m z - p \theta)} + \mathrm{c.c.},~~~~~\label{eq:eq18}\\
					v_z (r, \theta , z, t) &=& v_z^0 J_p (l r)  e^{i (\omega t - m z - p \theta)} + \mathrm{c.c.},~~~~~\label{eq:eq19}\\
					b (r,\theta,z,t) &=& i\frac{N^2 v_z^0}{\omega} J_p (l r)  e^{i (\omega t - m z - p \theta)} + \mathrm{c.c.},~~~~~\label{eq:eq20}\\
					P (r,\theta,z,t) &=& = \frac{\rho_0 m (f^2 - \omega^2)}{l^2 \omega} J_p (l r)  e^{i (\omega t - m z - p \theta)} + \mathrm{c.c.},~~~~~\label{eq:eq21}
				\end{eqnarray}
				with $J_p$ a Bessel function of order $p\in\mathbb{N}$, and with $\omega$ the wave frequency, and $l$, $m$, and $p$ the radial, vertical, and azimuthal wave numbers, respectively. Note that $l\in\mathbb{R}$ and $m\in\mathbb{R}$ have the dimension of spatial wave numbers ($\mathrm{m^{-1}}$), whereas $p\in\mathbb{Z}$ has the dimension of an angular wave number ($\mathrm{rad^{-1}}$). Other radial functions also satisfy equations~\eqref{eq:eq17}, \eqref{eq:eq18}, and~\eqref{eq:eq19}, but have divergences either for $r\rightarrow0$ or $r\rightarrow+\infty$ \citep{NIST2010}, and are therefore not considered in this problem. A more thorough description of these modes can be found in \cite{guimbardPhD, bouryPhD}.
		
\begin{figure}
				\centering
				\includegraphics[scale=1]{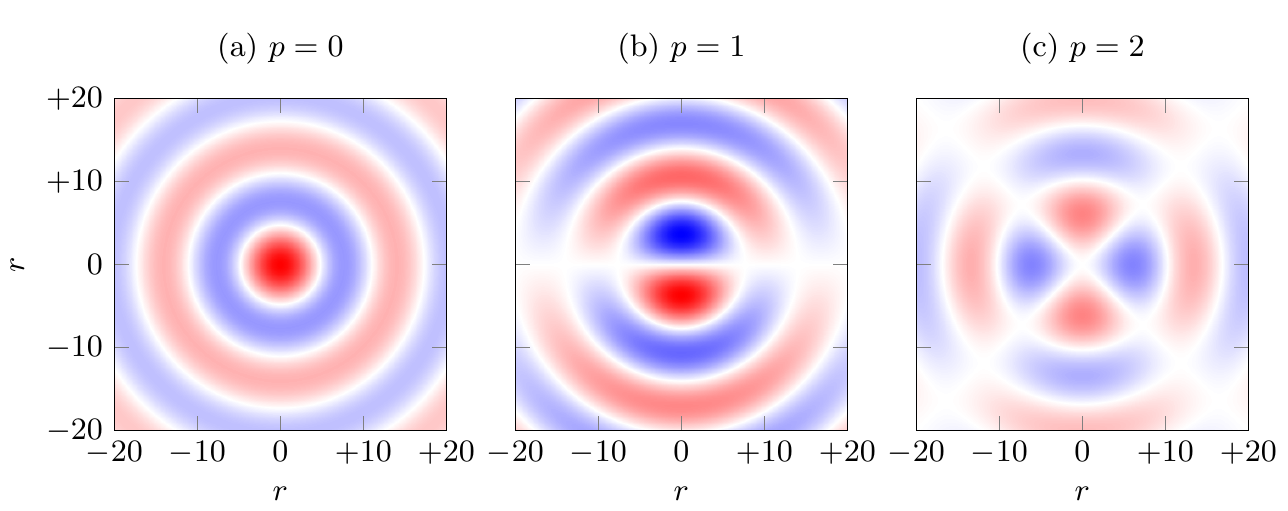}
				\caption{Vertical velocity field $v_z$ in a horizontal cross-section, for $p=0$ through $2$. For purpose of visualisation in the horizontal projection, the (dimensionless) radius $r$ is algebrical. The range of the colormap is purely indicative.}
				\label{fig:fig1}
			\end{figure}

			Figure~\ref{fig:fig1} presents plots of the vertical velocity field in the horizontal cross-section for different Kelvin modes as a function of dimensionless radius $r$, and whose horizontal structures correspond to $p=0$, $1$, and $2$, from left to right, respectively. The axisymmetric configuration $p=0$ (figure~\ref{fig:fig1}(a)) shows non-zero vertical velocity at the center of the domain $r=0$ and is $\theta$-invariant; conversely, modes at $p\neq 0$ have $v_z=0$ at $r=0$, and are $p$-periodic in $\theta$.
			
			In the simplified case of non-rotating flows ($f=0$) that will be considered from now on, and using relationships between Bessel functions, the system~\eqref{eq:eq17}--\eqref{eq:eq21} becomes
			\begin{eqnarray}
				v_r (r, \theta , z, t) &=& - i \frac{m v_z^0}{l} J_{p}' (l r) e^{i (\omega t - m z - p \theta)} + \mathrm{c.c.},\label{eq:pol1}\\
				v_\theta (r, \theta , z, t) &=& - \frac{m p v_z^0}{l^2 r} J_{p} (l r) e^{i (\omega t - m z - p \theta)} + \mathrm{c.c.},\label{eq:pol2}\\
				v_z (r, \theta , z, t) &=& v_z^0 J_p (l r)  e^{i (\omega t - m z - p \theta)} + \mathrm{c.c.},\label{eq:pol3}\\
				b (r,\theta,z,t) &=&  i\frac{N^2 v_z^0}{\omega} J_p (l r)  e^{i (\omega t - m z - p \theta)} + \mathrm{c.c.},\label{eq:pol4}\\
				P (r,\theta,z,t) &=& \frac{\rho_0 (N^2 - \omega^2) v_z^0}{\omega m} J_p (l r)  e^{i (\omega t - m z - p \theta)} + \mathrm{c.c.},\label{eq:pol5}
				\end{eqnarray}
				with $J_p$ a Bessel function of order $p\in\mathbb{N}$ and $J_p'$ its first derivative. These are consistent with the axisymmetric case $p=0$ discussed, for example, by \cite{ansong2010} and by \cite{boury2018}. Note that, for axisymmetric wave fields, $v_\theta$ is automatically zero since $v_\theta \propto p$, whereas it is not true for cylindrical wave fields. We then deduce the polarisation relations that give the velocity, buoyancy, and pressure fields as a function of the vertical velocity $v_z$ as
				\begin{equation}
					v_r = -i\frac{m}{l^2} \partial_r v_z, \mathrm{~~~} v_\theta = -\frac{mp}{l^2 r} v_z,\mathrm{~~~} b = i\frac{N^2}{\omega} v_z, \mathrm{~~~and~~~} P = \frac{\rho_0 (N^2 - \omega^2)}{\omega m} v_z.
				\end{equation}

		\subsection{Non-Linearities and Sub-harmonics Generation}
	
			In the present study, we focus on three-wave interactions constituting internal wave triads. To provide more insights on this phenomenon, we discuss here the fully non-linear equations and compute explicitly the non-linear terms in the case of three monochromatic Kelvin modes. Theoretical derivations are now conducted in the non-rotating case, in order to make the calculus more tractable and pedagogical; note that the subsequent discussion is not changed if we consider rotating flows. Setting $f=0$, the system of equations \eqref{eq:eq1} -- \eqref{eq:eq3} is equivalent, once expanded, to
			\begin{eqnarray}
				\partial_t v_r + \frac{1}{\rho_0} \partial_r P &=& - (\mathbf{v} \cdot \mathbf{\nabla}) v_r, \label{eq3.25}\\
				\partial_t v_\theta + \frac{1}{r \rho_0}\partial_\theta P  &=& - (\mathbf{v} \cdot \mathbf{\nabla}) v_\theta, \label{eq3.26}\\
				\partial_t v_z + \frac{1}{\rho_0}\partial_z P - b &=& - (\mathbf{v} \cdot \mathbf{\nabla}) v_z, \label{eq3.27}\\
				\partial_t b + N^2 v_z &=& - (\mathbf{v} \cdot \mathbf{\nabla}) b, \label{eq3.28}\\
				\mathbf{\nabla}\cdot\mathbf{v} &=& 0. \label{eq3.29}
			\end{eqnarray}	
			In the general case, the triadic wave field can be decomposed into
			\begin{equation}
				v_r = \sum_{j=1}^3 v_{r,j}, \mathrm{~~~} v_\theta = \sum_{j=1}^3 v_{\theta,j}, \mathrm{~~~} v_z = \sum_{j=1}^3 v_{z,j}, \mathrm{~~~} b = \sum_{j=1}^3 b_{j}, \mathrm{~~~and~~~} P = \sum_{j=1}^3 P_{j},\label{eq:eqsum3.33}
			\end{equation}
			where a canonical Kelvin mode, labeled $j$, is defined according to equations~\eqref{eq:pol1}--\eqref{eq:pol5} as
			\begin{eqnarray}
				v_{r,j} (r,\theta,z,t) &=& v_{r,j}^0(t) J_{p_j}' (l_j r) e^{i(\omega_j t - m_j z - p_j \theta)} + \mathrm{c.c.}, \label{eq3.31}\\
				v_{\theta,j} (r,\theta,z,t) &=& v_{\theta,j}^0(t) \frac{J_{p_j} (l_j r)}{l_j r} e^{i(\omega_j t - m_j z - p_j \theta)} + \mathrm{c.c.}, \label{eq3.32}\\
				v_{z,j} (r,\theta,z,t) &=& v_{z,j}^0(t) J_{p_j} (l_j r) e^{i(\omega_j t - m_j z - p_j \theta)} + \mathrm{c.c.}, \label{eq3.33}\\
				b_j (r,\theta,z,t) &=& b_{j}^0(t) J_{p_j} (l_j r) e^{i(\omega_j t - m_j z - p_j \theta)} + \mathrm{c.c.}, \label{eq3.34}\\
				P_j (r,\theta,z,t) &=& P_{j}^0(t) J_{p_j} (l_j r) e^{i(\omega_j t - m_j z - p_j \theta)} + \mathrm{c.c.}, \label{eq3.35}
			\end{eqnarray}
			where $\omega_j$ is a frequency, and $l_j$, $m_j$, and $p_j$ are wave numbers. In order to solve the system and allow energetic exchanges between the modes, the amplitudes $v_{r,j}^0$, $v_{\theta,j}^0$, $v_{z,j}^0$, $b_j^0$, and $P_j^0$ are slowly varying in time, yet still uniform in space. The radial dependence, involving Bessel derivative and $1/r$ prefactor, comes from the linear solution derived in the previous subsection.

			 The vectorial equation~\eqref{eq3.25}--\eqref{eq3.28} and the scalar equation~\eqref{eq3.29} constitute a system with four non-linear equations where, introducing the triad~\eqref{eq:eqsum3.33} and the explicit writing of the fields~\eqref{eq3.31}--\eqref{eq3.35}, the linear left-hand side can be written
			\begin{eqnarray}
				\partial_t v_r + \frac{1}{\rho_0} \partial_r P &=& \sum_{j=1}^3 \left[\partial_t v_{r,j}^0 +  i \omega_j v_{r,j}^0 + \frac{l_j}{\rho_0} P_j^0 \right] J_{p_j}'(l_j r) e^{i(\omega_j t - m_j z - p_j \theta)} + \mathrm{c.c.}, \label{eq3.36}
				\\
				\partial_t v_\theta + \frac{1}{r \rho_0}\partial_\theta P  &=& \sum_{j=1}^3 \left[\partial_t v_{\theta,j}^0 + i \omega_j v_{\theta,j}^0 - i \frac{p_j l_j}{\rho_0} P_j^0 \right] \frac{J_{p_j} (l_j r)}{l_j r} e^{i(\omega_j t - m_j z - p_j \theta)} + \mathrm{c.c.}, \label{eq3.37}
				\\
				\partial_t v_z + \frac{1}{\rho_0}\partial_z P - b &=& \sum_{i=j}^3 \left[\partial_t v_{z,j}^0 + i\omega_j v_{z,j}^0 - i \frac{m_j}{\rho_0} P_j^0 \right]  J_{p_j} (l_j r) e^{i(\omega_j t - m_j z - p_j \theta)} + \mathrm{c.c.}, \label{eq3.38}
				\\
				\partial_t b + N^2 v_z &=& \sum_{i=j}^3 \left[ \partial_t b_{j}^0 + i\omega_j b_j^0 + N^2 v_{z,j}^0 \right] J_{p_j} (l_j r) e^{i(\omega_j t - m_j z - p_j \theta)} + \mathrm{c.c.}. \label{eq3.39}
			\end{eqnarray}
			At this point, it is worth noting that equations \eqref{eq3.36}--\eqref{eq3.39} show a well established structure for the radial dependence (namely \eqref{eq3.36} $\propto J_{p_j}'(r)$, \eqref{eq3.37} $\propto J_{p_j} (r) / r$, and \eqref{eq3.38} and \eqref{eq3.39} $\propto J_{p_j}(r)$). In order to simplify these equations, and to be able to compute scalar products involving Bessel functions (see next section), it is better to write the radial dependences from equations \eqref{eq3.36} and \eqref{eq3.37} only in terms of a single Bessel function $\propto J_{p_j}(r)$ as in equations \eqref{eq3.38} and \eqref{eq3.39}. This can be done easily by integrating \eqref{eq3.36}, and by multiplying \eqref{eq3.37} by $r$; as a result, the linear terms are equivalent to
			\begin{eqnarray}
				\int \left( \partial_t v_r + \frac{1}{\rho_0} \partial_r P \right) \diff r &=& \sum_{j=1}^3 \left[\partial_t v_{r,j}^0 + i \frac{\omega_j}{l_j} v_{r,j}^0 + \frac{1}{\rho_0} P_j^0 \right] J_{p_j}(l_j r) e^{i(\omega_j t - m_j z - p_j \theta)} + \mathrm{c.c.}, \label{eq3.40}
				\\
				r \left(\partial_t v_\theta + \frac{1}{r \rho_0}\partial_\theta P\right)  &=& \sum_{j=1}^3 \left[\partial_t v_{\theta,j}^0 + i \frac{\omega_j}{l_j} v_{\theta,j}^0 - i \frac{p_j}{\rho_0} P_j^0 \right] J_{p_j} (l_j r) e^{i(\omega_j t - m_j z - p_j \theta)} + \mathrm{c.c.}, \label{eq3.41}
				\\
				\partial_t v_z + \frac{1}{\rho_0}\partial_z P - b &=& \sum_{i=j}^3 \left[\partial_t v_{z,j}^0 +  i\omega_j v_{z,j}^0 - i \frac{m_j}{\rho_0} P_j^0 \right]  J_{p_j} (l_j r) e^{i(\omega_j t - m_j z - p_j \theta)} + \mathrm{c.c.}, \label{eq3.42}
				\\
				\partial_t b + N^2 v_z &=& \sum_{i=j}^3 \left[ \partial_t b_{j}^0 + i\omega_j b_j^0 + N^2 v_{z,j}^0 \right] J_{p_j} (l_j r) e^{i(\omega_j t - m_j z - p_j \theta)} + \mathrm{c.c.}. \label{eq3.43}
			\end{eqnarray}
			
			Following the same techniques that have been used to investigate Cartesian TRI, we consider a time scale separation in which the amplitude variations have a different temporal scale than the wave field itself, i.e. $\partial_t v_z^0 \ll \omega v_z^0 $. This means that, at first order, we recover the polarisation relations and the linear solution whereas at second order, we obtain equations on the amplitudes linked to the non-linear interaction terms of the right-hand sides of \eqref{eq3.25}--\eqref{eq3.29} that are already of second order or more. In addition, this scale separation imposes that the amplitudes of all fields have the same temporal variation, i.e. $\partial_t v_{z,j}^0 \propto \partial_t v_{r,j}^0\propto \partial_t v_{\theta,j}^0\propto \partial_t b_{j}^0$. The study of the non-linear terms can therefore be reduced to the investigation of the effect of the non-linear terms on $v_{z,j}^0$. We now focus on these non-linear terms, to which we apply the same operations as on the linear terms (respectively integrating with respect to $r$ the radial equation on $v_r$, multiplying by $r$ the orthoradial equation on $v_\theta$, and letting unchanged the equations on $v_z$ and $b$). In order to compute the non-linear terms, we consider a triad of Kelvin modes, defined through the vertical velocity as follows
			\begin{equation}
				v_z (r,\theta,z,t) = \sum_{i=1}^3 v_{z,i}(r,\theta,z,t) = \sum_{i=1}^3 v_{z,i}^0 (t) J_{p_i} (l_i r) e^{i(\omega_i t - m_i z - p_i \theta)} + \mathrm{c.c.}. \label{eq3.44}
			\end{equation}
			As discussed previously, using the polarisation relations \eqref{eq:pol1} through \eqref{eq:pol5}, the choice of vertical velocity is sufficient to describe the whole velocity and buyoancy fields. As detailed in appendix~\ref{app:nonlinearterms}, the computation of the non-linear terms shows that the behaviour of $(\mathbf{v} \cdot \mathbf{\nabla}) v_z$ determines the structure of the sub-harmonics, and we will therefore restrict our study to this term. We will write
			\begin{equation}
				- (\mathbf{v} \cdot \mathbf{\nabla}) v_z = \sum_{i=1}^3 \sum_{j=1}^3 -i  \frac{v_{z,i}^0 v_{z,j}^0}{l_j} \left[\frac{M_{ij}}{2} \left( \mathsf{J}_{p_i -1}^{p_j+1} + \mathsf{J}_{p_i +1}^{p_j-1} \right) - M_{ji} \mathsf{J}_{p_i}^{p_j}\right] \Phi_{ij},
			\end{equation}
			where the azimuthal, vertical, and temporal dependences are included in
			\begin{equation}
				\Phi_{ij} \equiv \Phi_{ij} (\theta,z,t) = e^{i\left[(\omega_i \pm \omega_j) t - (m_i \pm m_j) z - (p_i \pm p_j) \theta\right]},
			\end{equation}
			and where we have defined the wave number product
			\begin{equation}
				M_{ij} = m_i l_j.
			\end{equation}
			For the sake of clarity, we droped the arguments $l_i r$ and $l_j r$ in the Bessel functions and we defined the radial-dependent quantity
			\begin{equation}
				\mathsf{J}_{p_i}^{p_j} \equiv \mathsf{J}_{p_i}^{p_j} (r) = J_{p_i}(l_i r) J_{p_j} (l_j r).
			\end{equation}
			
			To summarize, the linear left-hand side is a sum of three monochromatic waves, corresponding to a single frequency and a single spatial configuration, whereas the non-linear right-hand side is a sum of interacting waves. We note that, in general, these terms are non-zero, even for the self-interaction term. As a result, they can act as a second order forcing term on the linear part of the equations, similarly to what has been discussed in the appendix of \cite{boury2019a} and in \cite{boury2020a}. This is relevant to triad formation as, similarly to what has been described in $2$D geometries by \cite{joubaud2012, bourget2013}, and \cite{maurer2016}, two sub-harmonic wave fields can grow out of noise with energy input from a monochromatic forcing, leading to Triadic Resonant Interaction (TRI) or, if the sub-harmonics are of the same frequency, Parametric Sub-harmonic Instability (PSI). A more thorough discussion is provided in the next sections.

	\section{Unconfined Domains: Triadic Resonant Instability}
	\label{sec:unconfined}

			\begin{table}
				\centering
				\begin{tabular}{lll}
					\hline 
					~ & Accessible domain & Scalar product \\ 
					\hline 
					\multirow{2}{*}{Temporal ($t$)} & $t \in ~]-\infty;\,+\infty[$ & \multirow{2}{*}{$\displaystyle \left\langle \left. v_{z,i} ~\right\vert~ v_{z,j} \right\rangle_t = \frac{1}{2\pi} \int_{-\infty}^{+\infty} e^{i (\omega_i-\omega_j) t} \diff t = \delta(\omega_i - \omega_j)$} \\ 
					& $\omega \in ~ [0;\,+\infty[$ &  \\
					\hline 
					\multirow{2}{*}{Radial ($r$)} & $r \in ~ [0;\,+\infty[$ & \multirow{2}{*}{$\displaystyle \left\langle \left. v_{z,i} ~\right\vert~ v_{z,j} \right\rangle_r = \int_0^{+\infty} J_{p_i} (l_i r) J_{p_j} (l_j r) r \diff r = \frac{\delta(l_i - l_j)}{l_i}$} \\ 
					& $l \in ~ ]-\infty;\,+\infty[$ &  \\
					& ~ & $~\mathrm{for~}p_i=p_j$ \\
					\hline 
					\multirow{2}{*}{Vertical ($z$)} & $z \in ~ ]-\infty;\,+\infty [$ & \multirow{2}{*}{$\displaystyle \left\langle \left. v_{z,i} ~\right\vert~ v_{z,j} \right\rangle_z = \frac{1}{2\pi} \int_{-\infty}^{+\infty} e^{-i (m_i-m_j) z} \diff z = \delta(m_i - m_j)$} \\ 
					& $m \in ~ ]-\infty;\,+\infty[$ &  \\
					\hline 
					\multirow{2}{*}{Azimuthal ($\theta$)} & $\theta \in ~ [0;\,2\pi[$ & \multirow{2}{*}{$\displaystyle \left\langle \left. v_{z,i} ~\right\vert~ v_{z,j} \right\rangle_\theta = \frac{1}{2\pi} \int_0^{2\pi} e^{-i (p_i-p_j) \theta} \diff \theta = \delta(p_i - p_j)$} \\ 
					& $p \in\mathbb{Z}$ &  \\
					\hline 
				\end{tabular}
				\caption{Scalar products to consider in an unconfined domain.} 
				\label{tab:unconfined}
			\end{table}
	
		\subsection{Physical Domain and Projection}
			
			In order to investigate the forcing term created by the non-linear interactions, we generalise the projection method used to derive resonance relations in two-dimensional Cartesian TRI \citep{bourgetPhD, maurerPhD}. We have shown previously that, in the linear theory, Kelvin modes (i.e. the velocity, buoyancy, and pressure fields solutions of the linear equations) can be entirely determined by the vertical velocity field, $v_z$. Given different $4$-uplets $\left\lbrace \omega_i, l_i, p_i, m_i \right\rbrace$ of frequencies and wave numbers, the corresponding vertical velocity fields $\left\lbrace v_{z,i} \right\rbrace$ (defined by equation~\eqref{eq3.44}) form a familly of mutually orthogonal functions with the spatio-temporal Fourier-Hankel scalar product defined in table~\ref{tab:unconfined} through mathematical identities on exponentials and Bessel functions. The complete scalar product (operated for $(t,r,\theta,z) \in \mathbb{R} \times \mathbb{R^+} \times [0;\,2\pi] \times \mathbb{R}$) of two fields will be noted $\left\langle v_{z,i} | v_{z,j} \right\rangle$.

			Here, we should point out an additional difficulty compared to the Cartesian case: the projections are different when one focuses either on the temporal and vertical variables (integrals over $\mathbb{R}$), on the azimuthal variable (integral on $[0;\,2\pi]$ due to the $2\pi$ periodicity), or on the radial coordinate (integral on $\mathbb{R}^+$, with orthogonality of Bessel functions). Note that the previous re-writing of the system of governing equations \eqref{eq3.40}--\eqref{eq3.43}, with the multiplication by $r$ and the integration, is fully justified by the structure of radial scalar product of the $v_r$, $v_\theta$, and $v_z$ equations, since we need them to depend only on Bessel functions (and, notably,  not on derivatives). As previously detailed, we only consider equation~\eqref{eq3.27} on $v_z$, and its projection on a monochromatic solution defined by equation~\eqref{eq3.33}, of norm $1$, noted $v_{z,j}^\ast$
			\begin{equation}
				v_{z,j}^\ast (r,\theta,z,t) = J_{p_j} (l_j r) e^{i(\omega_j t - m_j z - p_j \theta)},
			\end{equation}
			leads to
			\begin{equation}
				\left\langle \left. \partial_t v_z + \frac{1}{\rho_0}\partial_z P - b~\right\vert~ v_{z,j}^\ast \right\rangle = \partial_t v_{z,j}^0 = \left\langle \left. - \left( \mathbf{v}\cdot\mathbf{\nabla} \right) v_z ~\right\vert~ v_{z,j}^\ast \right\rangle,\label{eq:scalar2}
			\end{equation}
			Since the temporal variations of the amplitudes present in the linear terms are at a different time scale than the oscillatory part, they appear as uncoupled and the scalar product does not affect their derivatives. We conclude, from equation~\eqref{eq:scalar2}, that the slow-varying amplitude terms can be fed by non-linear processes if and only if the scalar product of the corresponding non-linear left-hand side is non-zero.

			\subsection{Frequency Resonance}
			
				Performing the temporal scalar product on the non-linear terms, we obtain a delta function whose argument is a linear combination of three frequencies. As the frequencies are non-zero, the only resonant term, i.e. non-zero, is obtained when the following resonance condition in frequency is satisfied
				\begin{equation}
					\omega_0 = \pm \omega_1 \pm \omega_2.
				\end{equation}
				In the present study, we focus on the generation of sub-harmonics, meaning that both $\omega_1$ and $\omega_2$ are smaller than $\omega_0$, but the existence of triads involving super-harmonics are also allowed by this relation. Note that a particular case exists when $\omega_1 = \omega_2 = \omega_0 / 2$, called Parametric Subharmonic Instability (PSI), with several experimental and in-situ oceanic observations.
			
			\subsection{Vertical and Azimuthal Resonances}
			
				Similarly, the vertical and azimuthal scalar products yield the following resonance conditions in vertical and azimuthal wave numbers
				\begin{eqnarray}
					m_0 &=& \pm m_1 \pm m_2,\\
					p_0 &=& \pm p_1 \pm p_2.
				\end{eqnarray}
				There is, however, a difference between the wave numbers $m$ and $p$, that will be of primary importance when comparing these resonances to the ones found in the confined domain: here, the vertical wave number $m$ is a continuous parameter that can take any value in $\mathbb{R}$, whereas the azimuthal wave number $p$ is, because of the $2\pi$-periodicity of the system, a discrete parameter taken in $\mathbb{Z}$.
				
			\subsection{Radial Resonance}
			
				The radial scalar products are more difficult to evaluate, as they involve integrals over a product of three Bessel functions of different orders and different arguments. For the purpose of the subsequent discussion, we define the following integral
				\begin{equation}
					\forall (h,i,j)\in \mathbb{N}^3,~\Xi_{hij} = \int _0 ^{+\infty} J_h (l_0 r) J_i (l_1 r) J_j (l_2 r) r \, \diff r. \label{eq4.13}
				\end{equation}
				
\begin{figure}
				\centering
				\includegraphics[scale=1]{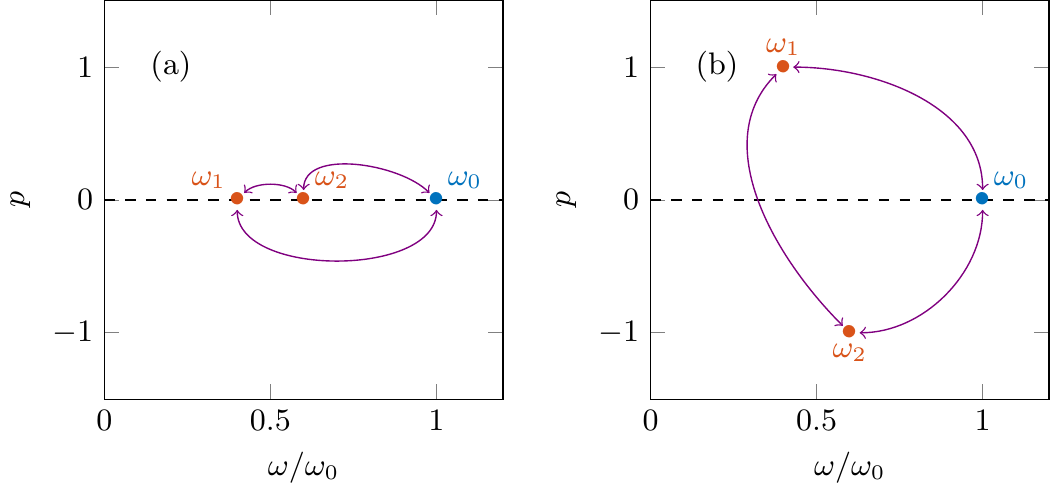}
				\caption{Two examples of interacting triads in the $(\omega,p)$-plane with (a) three axisymmetric modes and (b) an axisymmetric mode forcing forcing two modes in contra-rotation in the horizontal plane. The primary wave at $\omega_0$ (in blue) is forcing two sub-harmonics at $\omega_1$ and $\omega_2$ (in orange). The arrows show the interactions between the modes.}
				\label{fig:schematicradialexamples}
			\end{figure}

				Before trying to tackle the general case using asymptotics, we will delve into two peculiar examples, as depicted in figure~\ref{fig:schematicradialexamples}. The first case shown in figure~\ref{fig:schematicradialexamples}(a) is the triadic interaction of three axisymmetric modes; the second case presented in figure~\ref{fig:schematicradialexamples}(b) involves a symmetry breaking and leads to two cylindrical modes that are contra-rotating in the horizontal plane, forced by an axisymmetric primary wave. In both cases, one of the secondary (labeled $2$) is computed using the non-linear interaction of the primary wave (labeled $0$) with the other secondary wave (labeled $1$), which naturally introduces a non-symmetry between the two secondary waves; the calculus, however, could also be performed by switching the secondary waves $1$ and $2$ and would lead to similar results. 
	
			\subsubsection{Radial Resonance: Axisymmetric Case $p_0 = p_1 = p_2 = 0$}
			
				We first consider the particular case of a triad constituted of three axisymmetric wave fields, i.e. with $p_0 = p_1 = p_2 = 0$. In such a triad, the axisymmetric primary wave is in resonance with two axisymmetric secondary waves (figure~\ref{fig:schematicradialexamples}(a)). The radial scalar products of the non-linear term writes
				\begin{equation}
					\left\langle- (\mathbf{v} \cdot \mathbf{\nabla}) v_z\right\rangle_r = A_{z} \left[ M_{01} \Xi_{110} + M_{10} \Xi_{000} \right], \label{eq4.16}
				\end{equation}
				with $A_z$ a constant coefficient. Note that this system is very similar to Cartesian non-linear systems \citep{bourgetPhD}. Equation~\eqref{eq4.16} is then resonant if and only if the right-hand side is non-zero.
			
			\begin{figure}
				\centering
				\includegraphics[scale=1]{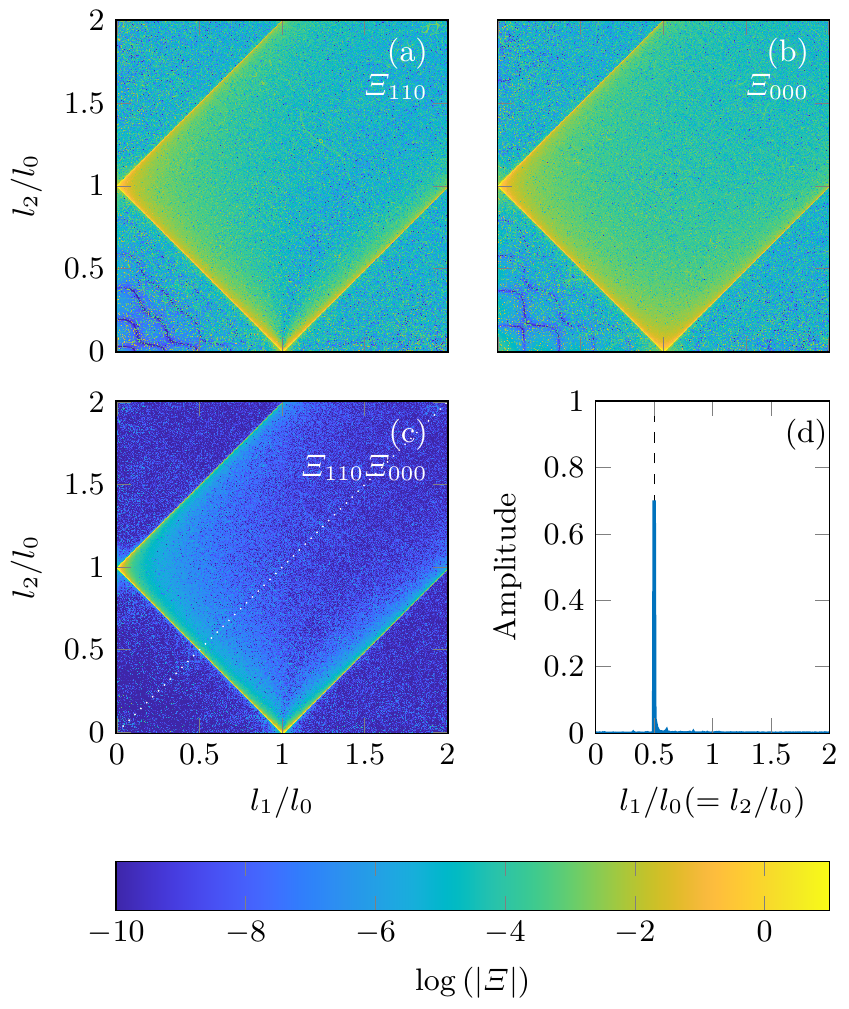}
				\caption{Colormaps of the logarithm of the coefficients $|\Xi_{hij}|$ as a function of wave number ratios $l_1/l_0$ and $l_2/l_0$, with: (a) $\log\left(|\Xi_{110}|\right)$, (b) $\log\left(|\Xi_{000}|\right)$, and (c) logarithm of the product of these two coefficients. Dashed lines show the locations of $l_0 \pm l_1 \pm l_2 = 0$. Plot (d) is the profile of $|\Xi_{110}\Xi_{000}|$ along the first bissectrix (show by the dotted line in plot (c)). }
				\label{fig:cmaps1}
			\end{figure}
				
				The coefficients $\Xi_{hij}$ involed in equation~\eqref{eq4.16} can be numerically investigated. Figure~\ref{fig:cmaps1} presents colormaps of the logarithm of the absolute value of these coefficients $|\Xi_{hij}|$ for: (a) $|\Xi_{110}|$; (b) $|\Xi_{000}|$; and (c) the product $|\Xi_{110}\Xi_{000}|$. The numerical integration if performed for $R l_0$ between $0$ and $1900$. All quantities are plotted as a function of $l_1/l_0$ and $l_2/l_0$, with $l_1/l_0$ and $l_2/l_0$ going from $0$ to $2$, and the colorbar saturates at $1$. The plots can be extended by symmetry and one can get the complete diagram for values of $l_1/l_0$ and $l_2/l_0$ that can be negative. Note that these quadrants are, in general, not symmetrical in respect with the bissectrix: switching the wave numbers $l_1$ and $l_2$ yields the same plot if and only if the corresponding indices of the associated Bessel functions in $\Xi_{hij}$ are the same i.e. if we can write $\Xi_{hhj}$ as in figures~\ref{fig:cmaps1}(a) and (b) with $\Xi_{000}$ and $\Xi_{110}$. As clearly identified in figure~\ref{fig:cmaps1}(c) (and less clearly in (a) and (b)), the only cases for which the $\Xi_{hij}$ integrals are non-zero correspond to the lines $l_0 \pm l_1 \pm l_2 = 0$, i.e. along the possible radial resonance relations. The plot in figure~\ref{fig:cmaps1}(d) show the value of $|\Xi_{000}\Xi_{110}|$ along the first bissectrix (dotted line in figure~\ref{fig:cmaps1}(c)) and illustrates that a maximum is reached when the resonance relation is reached, i.e. $l_1 / l_0 = l_2/l_0 = 0.5$.

			\subsubsection{Radial Resonance: Non-Axisymmetric Case $p_0 = 0$, $p_1=1$, and $p_2=-1$}
			
				We now consider a second case study involving a symmetry breaking, with $p_0 = 0$, $p_1=1$, and $p_2=-1$, corresponding to figure~\ref{fig:schematicradialexamples}(b). This triad corresponds to an axisymmetric primary wave in resonance with two non-axisymmetric (i.e. cylindrical) secondary waves. In the horizontal plane, one of the secondary waves is rotating clockwise while the other one is rotating anti-clockwise. Following the same reasoning as previously, the radial scalar product of the non-linear term writes
				\begin{equation}
					\left\langle- (\mathbf{v} \cdot \mathbf{\nabla}) v_z\right\rangle_r = A_{z} \left[ \frac{ M_{01}}{2}( \Xi_{101} - \Xi_{121}) - M_{10} \Xi_{011} \right]. \label{eq4.20}
				\end{equation}
			
			\begin{figure}
				\centering
				\includegraphics[scale=1]{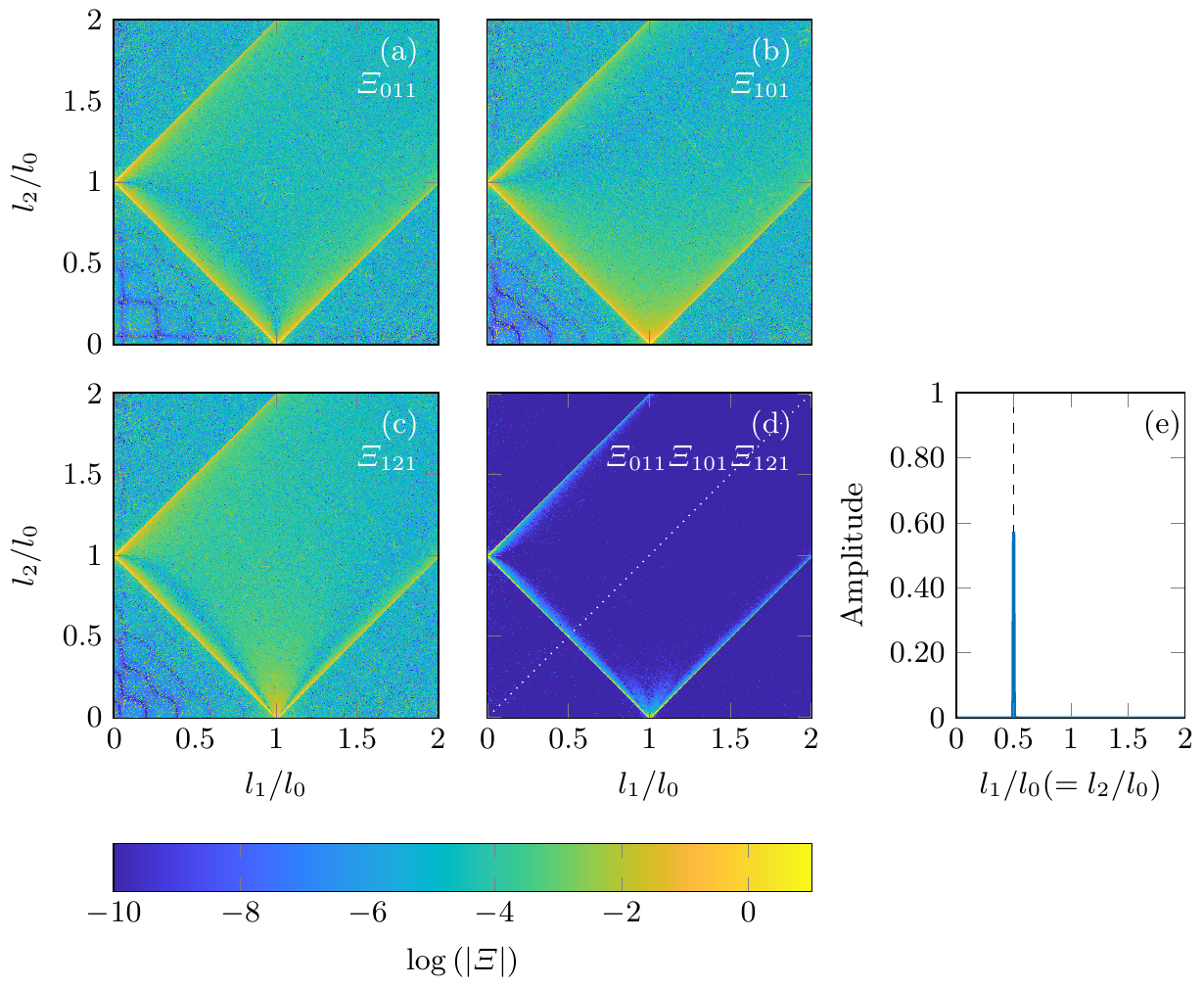}
				\caption{Colormaps of the logarithm of the coefficients $|\Xi_{hij}|$ as a function of wave number ratios $l_1/l_0$ and $l_2/l_0$, with: (a) $\log\left(|\Xi_{011}|\right)$, (b) $\log\left(|\Xi_{101}|\right)$, (c) $\log\left(|\Xi_{121}|\right)$, and (d) logarithm of the product of the three coefficients. Dashed lines show the locations of $l_0 \pm l_1 \pm l_2 = 0$. Plot (e) shows the profile of the product $|\Xi_{011}\Xi_{101}\Xi_{121}|$ along the first bissectrix (corresponding to the dotted line in plot (d)). }
				\label{fig:cmaps2}
			\end{figure}
			
			As for the previous case study, we conduct a numerical investigation of the different terms $|\Xi_{hij}|$ involved in equation~\eqref{eq4.20}. Figure~\ref{fig:cmaps2} presents colormaps of the logarithm of the absolute value of these coefficients for: (a) $|\Xi_{011}|$; (b) $|\Xi_{101}|$; (c) $|\Xi_{121}|$; and (d) the product $|\Xi_{011}\Xi_{101}\Xi_{121}|$. Again, the numerical integration is performed for $R l_0$ between $0$ and $1900$, and all quantities are plotted as a function of $l_1/l_0$ and $l_2/l_0$, with $l_1/l_0$ and $l_2/l_0$ going from $0$ to $2$. As in the fully axisymmetric case, the coefficients (and their product) are maximal along the lines corresponding to the radial resonance relation, and almost zero everywhere else.

			\subsubsection{Asymptotics}
			
\begin{figure}
				\centering
				\includegraphics[scale=1]{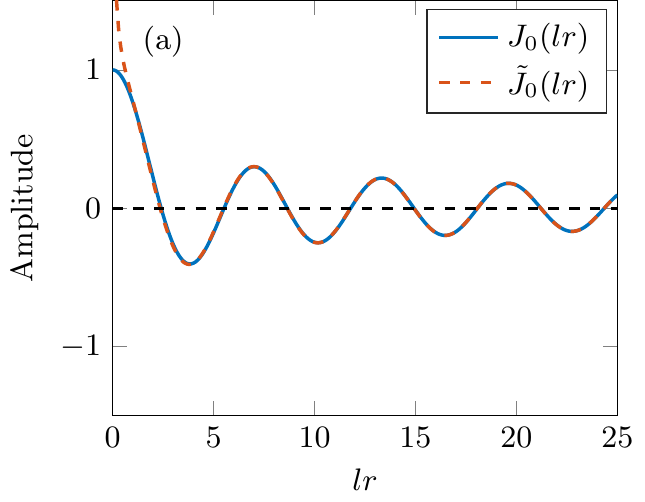}
				\includegraphics[scale=1]{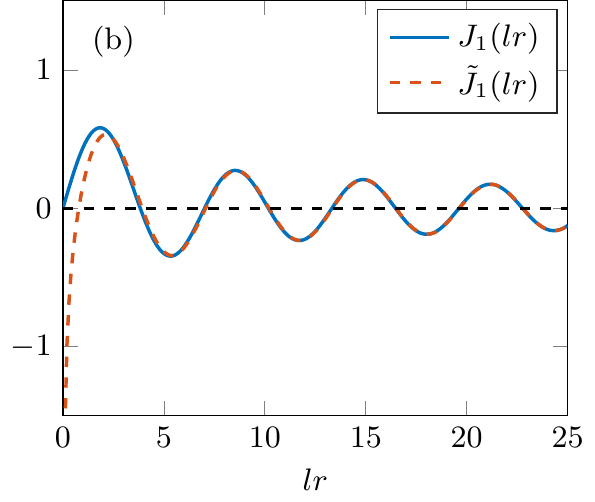}
				\caption{Plots of the functions (a) $J_0$ and (b) $J_1$ (solid lines) with the asymptotic approximation $\tilde{J}_0$ and $\tilde{J}_1$ (dashed lines).}
				\label{fig:asymptot}
			\end{figure}

				From the two numerical case studies with axisymmetric (figure~\ref{fig:schematicradialexamples}(a), case $1$) and cylindrical (figure~\ref{fig:schematicradialexamples}(b), case $2$) sub-harmonics, we empirically conjecture that the existence of non-vanishing $\Xi_{hij}$ coefficients in the non-linear part of the wave equations leads to a radial resonance condition of the form $l_0 = \pm l_1 \pm l_2$. A possible way to investigate this further is to use an asymptotic development of the Bessel functions. At a given radial wave number $l$, for large values of $l r$, the functions $J_n$ can be approximated by the functions $\tilde{J}_n$ defined as follows~\citep{NIST2010}
		\begin{equation}
			\forall n \in \mathbb{N},~\forall r \in \mathbb{R}^*,~\tilde{J}_n(lr) = \sqrt{\frac{2}{\pi lr}} \cos \left( lr -\frac{\pi}{2}\left(n+\frac{1}{2}\right) \right),\label{chap7:eq114}
		\end{equation}
		from which we deduce, notably, for the zeroth and first order Bessel functions describing the axisymmetric wave field
		\begin{equation}
			\forall r \in \mathbb{R}^*,~\tilde{J}_0 (lr) = \sqrt{\frac{2}{\pi l r}} \cos \left(lr-\frac{\pi}{4}\right)~~~~~~~\mathrm{and}~~~~~~\tilde{J}_1 (lr) = \sqrt{\frac{2}{\pi l r}} \sin \left(lr-\frac{\pi}{4}\right).
		\end{equation}
		These approximations are presented in figure~\ref{fig:asymptot} (a) and (b), respectively. As can be seen in these two plots, the values of $l r$ for which the approximation~\eqref{chap7:eq114} is valid (less than $4$\% difference) is $l r>1$ for $J_0$ and $l r> 2$ for $J_1$. Compared to our experimental configuration of a radial mode $1$ with $l=19\mathrm{~m^{-1}}$, this means that the profile is very well approximated by this decaying cosine for $r>15\mathrm{~cm}$, and sooner for higher order radial modes.
			
			The coefficients $\Xi_{hij}$ previously defined in equation~\eqref{eq4.13} can be rewritten, using this asymptotic formulation, as an improper integral
			\begin{equation}
				\forall (h,i,j)\in \mathbb{N}^3,~\Xi_{hij} \simeq \lim_{\varepsilon\rightarrow 0} \int _\varepsilon ^{+\infty} \tilde{J}_h (l_0 r) \tilde{J}_i (l_1 r) \tilde{J}_j (l_2 r) r \, \diff r, \label{chap7:eq113}
			\end{equation}
			whose integrand are diverging in $0$ (see dotted curves in figure~\ref{fig:asymptot}), while remaining integrable (which is why, formally, the limit $\varepsilon$ approaching $0$ is needed). Using the definition from equation~\eqref{chap7:eq114} and trigonometric relations, we find that the coefficients $\Xi_{hij}$ can be expressed as a sum of integrals over approximated Bessel functions
			\begin{equation}
				\forall (h,i,j)\in \mathbb{N}^3,~\Xi_{hij} \simeq  \Gamma(1,h,i,j) + \Gamma(0,h,i,-j) + \Gamma (0,h,-i,j) + \Gamma(-1,h,-i,-j), \label{eq4.25}
			\end{equation}
			where, given $(a,b,c,d)\in\mathbb{Z}^4$, we write
			\begin{equation}
				\Gamma (a,b,c,d) = \lim_{\varepsilon\rightarrow 0}\int _\varepsilon ^{+\infty} \frac{1}{2\pi} \sqrt{\frac{l_{bcd}}{l_0 l_1 l_2}} {\tilde{J}_{a+b+c+d} (l_{bcd} r)} \,\diff r,
			\end{equation}
			with, using the $\mathrm{sign}$ function, the radial interaction wave number $l_{bcd}$ (that can be negative) defined by
			\begin{equation}
				l_{bcd} = \mathrm{sign(b)} l_0 + \mathrm{sign(c)} l_1 + \mathrm{sign(d)} l_2. \label{eq:lbcd}
			\end{equation}
			Note that, thanks to the symmetric writing of $\Gamma$, the four $\Gamma$ integrals involved in equation~\eqref{eq4.25} are linked to four different radial interaction wave number $l_{abc}$ as shown in table~\ref{chap7:tabLbcd}. Interestingly, the cases $l_{bcd}=0$ correspond to the four possible triads that can be obtained through the formula $l_0 \pm l_1 \pm l_2 = 0$.
		\begin{table}
			\centering
			\begin{tabular}{c c c c c}
				\hline
				$\Gamma (a,b,c,d)$ & $\Gamma (1,h,i,j)$ & $\Gamma (0,h,i,-j)$ & $\Gamma (0,h,-i,j)$ & $\Gamma (-1,h,-i,-j)$ \\
				\hline\hline
				$l_{bcd}$ & $l_0 + l_1 + l_2$ & $l_0 + l_1 - l_2$ & $l_0 - l_1 + l_2$ & $l_0 - l_1 - l_2$ \\
				\hline
			\end{tabular}
			\caption{Radial interaction wave number $l_{bcd}$ corresponding to the different $\Gamma$ integrals.}
			\label{chap7:tabLbcd}
		\end{table}
			
			Thanks to trigonometric relations and change of variables, these $\Gamma$ integrals can be explictly described by a sum of Fresnel integrals $x \mapsto \mathsf{C}(x)$ and $x \mapsto \mathsf{S}(x)$ (see appendix~\ref{sec:app:fresnel} and \cite{NIST2010}), whose values in $x=0$ are $1$ and $0$, respectively, and whose limit when $x$ approaches $+\infty$ is $0$. We deduce that the value of $\Gamma$ only depends on the cosine Fresnel integral and we can therefore write
			\begin{equation}
				\Gamma(a,b,c,d) = \Gamma^\mathsf{C} (a,b,c,d),
			\end{equation}
			with\begin{equation}
				\Gamma^\mathsf{C} (a,b,c,d) = \sqrt{\frac{2}{\pi^3 l_0 l_1 l_2}} \cos \left(\frac{\pi}{2}\left( a+b+c+d +\frac{1}{2}\right) \right)  \delta(l_{bcd}).
			\end{equation}
			
		\begin{table}
			\centering
			\begin{tabular}{c c c c c}
				\hline
				Triadic relations: If...  & $l_0 + l_1 + l_2 = 0$ & $l_0 + l_1 - l_2 = 0$ & $l_0 - l_1 + l_2 = 0$ & $l_0 - l_1 - l_2 = 0$ \\
				\hline \hline
				Then $l_0 + l_1 + l_2 = $ & $0$      & $ 2 l_2$ & $ 2 l_1$ & $ 2 l_0$ \\
				Then $l_0 + l_1 - l_2 = $ & $-2 l_2$ & $0$      & $ 2 l_0$ & $ 2 l_1$ \\
				Then $l_0 - l_1 + l_2 = $ & $-2 l_1$ & $ 2 l_0$ & $0$      & $ 2 l_2$ \\
				Then $l_0 - l_1 - l_2 = $ & $ 2 l_0$ & $-2 l_1$ & $-2 l_2$ & $0$      \\
				\hline
			\end{tabular}
			\caption{Disjunctive case study showing the values of $l_0 \pm l_1 \pm l_2$ when a triadic relation between the radial wave numbers is satisfied.}
			\label{chap7:tabLdisj}
		\end{table}
		
			If $l_0$, $l_1$, and $l_2$ are linked by a triadic relation so that $l_{bcd} = 0$ for given $(b,c,d)\in \mathbb{Z}^3$, then one (and only one) of the four $\Gamma$ integrals has a reduced interaction radial wave number $l_{bcd}$ equal to zero whereas the three other have non-zero reduced interaction radial wave numbers. The corresponding disjonctive case study is presented in table~\ref{chap7:tabLdisj}. We conclude that one (and only one) of the four $\Gamma$ integrals is non-zero, and we have
			\begin{equation}
				|\Xi_{hij}| = \frac{1}{\sqrt{\pi^3 l_0 l_1 l_2}} \mathrm{~~~~with~~~~~} l_0 \pm l_1 \pm l_2 = 0.
			\end{equation}
			This is true, for example, for $(hij) = (000)$, $(110)$, $(011)$, $(101)$, and $(121)$ (that correspond to $\Xi_{hij}$ involved in equations~\eqref{eq4.16} and \eqref{eq4.20}), consistent with the two case studies. The five $\Xi$ integrals are therefore non-zero, and have approximatively the same norm. It can also be shown that they are maximal since $\diff [C(x)/x] / \diff x = 0$ for $x=0$. In this case, when the three radial wave numbers are linked by a linear relation of the form $l_0 = \pm l_1 \pm l_2$, the non-linear system of internal wave equations reduces to a system that no longer involves neither $\Xi_{hij}$ nor Bessel integrals, allowing for the same resolution method as in Cartesian geometry. Although the result is not exact (since it is derived from asymptotic expressions of the Bessel functions), this is an interesting finding that may contribute to the derivation of the resonance relation.

		\subsection{Degrees of Freedom vs. Constraints}
			
			Let us now discuss the degrees of freedom of such a triadic interaction. The primary wave field being set, the triad is determined by the frequencies and wave numbers of the sub-harmonic secondary waves, which means $8$ parameters ($2\times 1$ frequencies and $2\times 3$ wave numbers). The constraints can be listed as follows: $2$ dispersion relations, and $4$ resonance conditions. Therefore, the system has two degrees of freedom, which also means that it has a degenerency in its solutions. The triad satisfies the following relations
			\begin{eqnarray}
				\omega_0 &=& \pm \omega_1 \pm \omega_2,\\
				m_0 &=& \pm m_1 \pm m_2,\\
				p_0 &=& \pm p_1 \pm p_2,\\
				l_0 &\simeq& \pm l_1 \pm l_2,
			\end{eqnarray}
			in which we remind that the approximate equality for the radial wave numbers $l_0$, $l_1$, and $l_2$, comes from the geometry itself and properties of the Bessel functions (see figures~\ref{fig:cmaps1} and \ref{fig:cmaps2}, showing a finite (but non-zero) width peak for the resonance condition).	Although, to our knowledge, no such observation has been reported, the triadic resonant relations should be similar for three-dimensional Cartesian wave fields. Note that, for (quasi) two-dimensional wave fields (i.e. Cartesian $2$D or axisymmetric), there are only $6$ free parameters ($2\times 1$ frequencies and $2\times 2$ wave numbers) for $5$ constraints ($2$ dispersion relations and $3$ resonance conditions), which means that the triadic system is mono-valuated and only admits a unique solution once one of the free parameters is fixed. For example, in $2$D, setting one of the sub-harmonic frequencies is enough to charaterise the whole wave field and the sub-harmonic wave numbers. Conversely, in $3$D, the frequency and one of the wave numbers of a sub-harmonic can be choosen independently in order to determine the whole wave field. 
		
	
			This analysis has been performed in the non-inertial case, i.e. with a Coriolis frequency $f=0$, but similar study can be undertaken with $f \neq 0$. We speculate that, for rotating flows, the previous coupling equations would be modified with additional cross-terms that would add more complexity to the non-linear resonant forcing, but our main results --namely, the resonance conditions-- would not be altered. In other words, since the base flow equations are identical in the rotating and in the non-rotating case, the resonance conditions will be the same but the involved coefficients, related to the growth rates, would be slightly modified. The important effects would appear on the selection of the modes (i.e. which modes would be the most unstable) rather than on the resonance per se. We also note that adding rotation increases the wave instability, the three dimensional effects, and is more likely to create symmetry breakings (such as, in our case, the creation of pure cylindrical modes with an azimuthal wave number $p \neq 0$ out of an axisymmetric forcing wave field) \citep{maurer2016,ha2021,mora2021}.

		\subsection{Experimental Observation}

		Experiments involving density stratified and rotating fluid, while generating axisymmetric inertia-gravity waves in an unconfined domain, are presented in~\cite{maurerPhD}. The primary aim of these experiments was to trigger TRI with high sensitivity in the regime $(\omega/f,\omega/N)$ in which it is the most likely to occur~\citep{maurer2016}. Incidently, some of these experiments have shown resonant triads in cylindrical geometry, with a symmetry breaking, as previously described in our study case $2$. This subsection focuses on the analysis of one of these experiments, run at buoyancy frequency $N=0.97\mathrm{~rad\cdot s^{-1}}$ and Coriolis frequency $f=0.294\mathrm{~rad\cdot s^{-1}}$. The forcing imposed at frequency $\omega_0 = 0.80\mathrm{~rad\cdot s ^{-1}}$ is a truncated Bessel function with a wave number $l_0=42\mathrm{~m^{-1}}$ and an amplitude $a=15\mathrm{~mm}$. 
		
\begin{figure}
			\centering
			\includegraphics[scale=1]{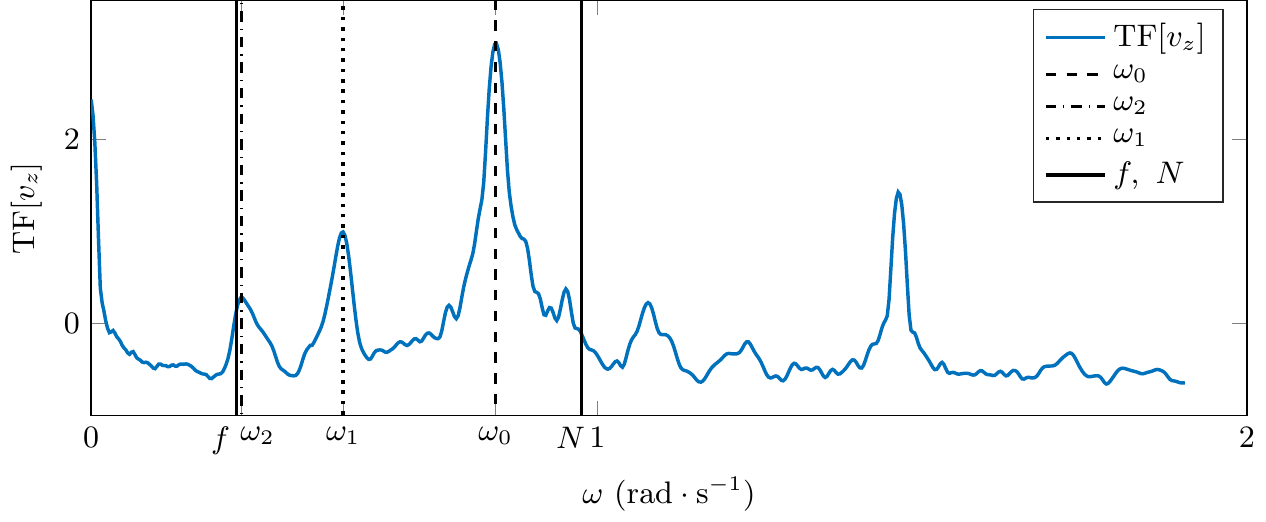}
			\caption{Fourier transform performed around $300\mathrm{~s}$ after starting the experiment. The buoyancy frequency is $N=0.97\mathrm{~rad\cdot s^{-1}}$ and the Coriolis frequency $f=0.294\mathrm{~rad\cdot s^{-1}}$. Forcing was imposed at a frequency $\omega_0 = 0.80\mathrm{~rad\cdot s ^{-1}}$.}
			\label{fig:spectrumpaco}
		\end{figure}

		On the Fourier transform computed around $300\mathrm{~s}$ after the beginning of the experiment, presented in figure~\ref{fig:spectrumpaco}, we can see a peak at the forcing frequency $\omega_0 = 0.80\mathrm{~rad\cdot s^{-1}}$ accompanied by two peaks at smaller frequencies, respectively $\omega_1 = 0.30\mathrm{~rad\cdot s^{-1}}$ and $\omega_2 = 0.50\mathrm{~rad\cdot s^{-1}}$. These three frequencies satisfy the triadic resonant condition $\omega_0 = \omega_1 + \omega_2$.
		
		
		
		\begin{figure}
			\centering
			\includegraphics[scale=1]{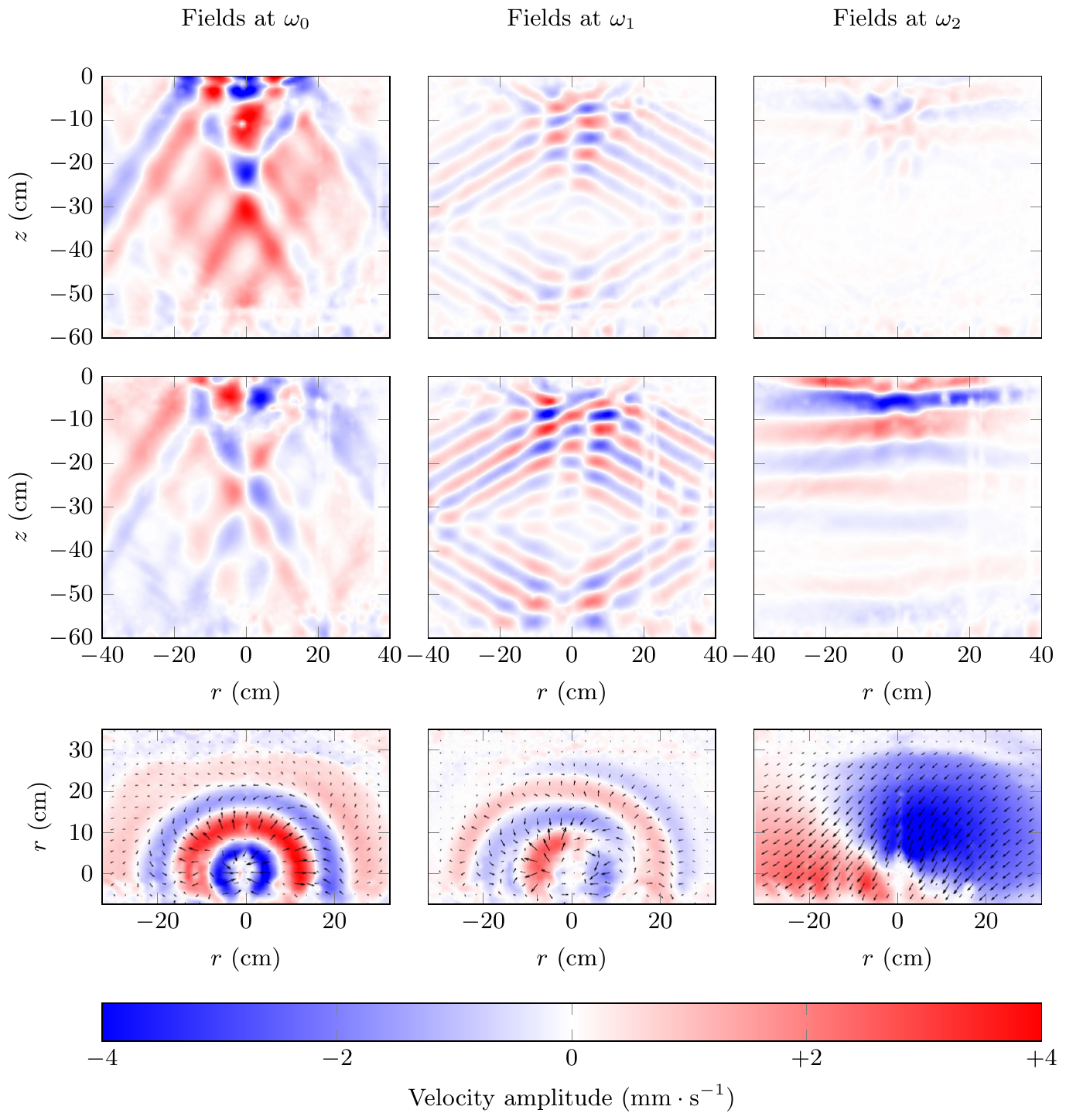}
			\caption{Radial velocity fields, with no confining cylinder, buoyancy frequency $N=0.97\mathrm{~rad\cdot s^{-1}}$ and Coriolis frequency $f=0.294\mathrm{~rad\cdot s^{-1}}$. From left to right: primary forcing wave at $\omega_0=0.80\mathrm{~rad\cdot s^{-1}}$, and secondary waves at $\omega_1=0.50\mathrm{~rad\cdot s^{-1}}$ and $\omega_2=0.30\mathrm{~rad\cdot s^{-1}}$. The first two rows show the vertical and the radial velocities in the vertical cross-section, and the third row shows the radial velocity in the horizontal cross-section. For the prupose of visualisation, $r$ is algebrical.}
			\label{fig:pacopiv}
		\end{figure}
		
		Velocity fields filtered at the frequencies associated with the observed TRI satisfying the resonance condition are presented in figure~\ref{fig:pacopiv}. In the vertical cross-section, we can estimate the vertical wave length and associated wave number $m_j$ for $j \in \left\lbrace 0,1,2 \right\rbrace$. These values are presented in table~\ref{chap7:tabMLP}. From our estimates, we verify the resonance condition on the vertical wave number as we have $m_0 \simeq m_1 - m_2$.		
		
		The radial wave fields are described by Bessel functions of the first kind $J_1$ and are therefore of the form $J_1 (l_j r)$ for $j \in \left\lbrace 0,1,2 \right\rbrace$. According to \cite{beattie1958}, the first zero of this Bessel function is equal to $3.83$. For each value of $j$, the location $r_j$ of the first zero can be identified in these velocity fields (figure~\ref{fig:pacopiv} bottom), and the corresponding wave number $l_j$ can then be deduced  ($l_j=3.83/r_j$). These numbers are presented in table~\ref{chap7:tabMLP} for the three frequencies identified in figure~\ref{fig:spectrumpaco}. The radial wave number $l_0=42\mathrm{~m^{-1}}$ obtained for the primary wave is consistent with the imposed forcing. The two radial wave numbers for the secondary waves are close to satisfy the resonance relation $l_0 \simeq l_1 + l_2$.
		
		Furthermore, there is a clear symmetry breaking as the velocity fields for the secondary waves at $\omega_1$ and $\omega_2$ start rotating clockwise and anti-clockwise, respectively, meaning that there is an azimuthal wave number $p_1= +1$ and $p_2=-1$ (see table~\ref{chap7:tabMLP}). This verifies the orthoradial resonance condition as the excitation field has an azimuthal wave number $p_0 = 0 = p_1 + p_2$. This is also consistent with the fact that the primary wave is axisymmetric, i.e. can be described by $v_z \propto J_0 (l_0 r)$ and $v_r \propto J_1 (l_0 r)$ with non-zero vertical velocity and zero radial velocity at $r=0$, whereas the two secondary waves are cylindrical and non-axisymmetric, for example with non-zero radial velocity at $r=0$, as allowed for Kelvin modes.
		
		\begin{table}
			\centering
			\begin{tabular}{l c c c}
				\hline
				Field at frequency $\omega_j$ & $\omega_0$ & $\omega_1$ & $\omega_2$ \\
				\hline\hline
				Vertical wave length $\mathrm{~(m)}$ & $0.40 \pm 0.04$ & $0.10 \pm 0.02$ & $0.13 \pm 0.02 $ \\
				Corresponding wave number $m_j\mathrm{~(m^{-1})}$ & $16 \pm 2$ & $63 \pm 8$ & $48 \pm 9$ \\ 
				\hline
				First radial zero identified $r_j\mathrm{~(m)}$ & $0.09 \pm 0.01$ & $0.12 \pm 0.01$ & $0.33 \pm 0.03 $ \\
				Corresponding wave number $l_j\mathrm{~(m^{-1})}$ & $42 \pm 5$ & $32 \pm 3$ & $12 \pm 1$ \\ 
				\hline
				Identified orthoradial periodicity $p_j$ & $0$ & $+1$ & $-1$\\
				\hline
			\end{tabular}
			\caption{Wave numbers extracted from the experiment for the radial velocity fields filtered at $\omega_j$, with $j\in \left\lbrace 0,1,2 \right\rbrace$, showing from top to bottom: the vertical wave lengths and their corresponding vertical wave numbers; the first zeros $r_j$ measured in the experiment and their corresponding radial wave number $l_j$; the identified orthoradial periodicity $p_j$.}
			\label{chap7:tabMLP}
		\end{table}

	\section{Confined Domains: Coercion by Boundary Conditions}
	\label{sec:confined}

			\begin{table}
				\centering
				\begin{tabular}{lll}
					\hline 
					~ & Accessible domain & Scalar product \\ 
					\hline 
					\multirow{2}{*}{Temporal ($t$)} & $t \in ~]-\infty;\,+\infty[$ & \multirow{2}{*}{$\displaystyle \left\langle \left. v_{z,i} ~\right\vert~ v_{z,j} \right\rangle_t = \frac{1}{2\pi} \int_{-\infty}^{+\infty} e^{i (\omega_i-\omega_j) t} \diff t = \delta(\omega_i - \omega_j)$} \\ 
					& $\omega \in ~ [0;\,+\infty[$ &  \\
					\hline 
					\multirow{2}{*}{Radial ($r$)} & $r \in ~ [0;\,R]$ & \multirow{2}{*}{$\displaystyle \left\langle \left. v_{z,i} ~\right\vert~ v_{z,j} \right\rangle_r = \int_0^{R} J_{p_i} (l_i r) J_{p_j} (l_j r) r \diff r = \frac{\delta(l_i - l_j)}{l_i}$} \\ 
					& $l R$ a Bessel zero &  \\
					& ~ & $~\mathrm{for~}p_i=p_j$ \\
					\hline 
					\multirow{2}{*}{Vertical ($z$)} & $z \in ~ [0;\,H]$ & \multirow{2}{*}{$\displaystyle \left\langle \left. v_{z,i} ~\right\vert~ v_{z,j} \right\rangle_z = \frac{1}{2\pi} \int_{0}^{H} e^{-i (m_i-m_j) z} \diff z = \delta(m_i - m_j)$} \\ 
					& $m = n\pi / (2 H), ~n\in \mathbb{Z}$ &  \\
					\hline 
					\multirow{2}{*}{Azimuthal ($\theta$)} & $\theta \in ~ [0;\,2\pi[$ & \multirow{2}{*}{$\displaystyle \left\langle \left. v_{z,i} ~\right\vert~ v_{z,j} \right\rangle_\theta = \frac{1}{2\pi} \int_0^{2\pi} e^{-i (p_i-p_j) \theta} \diff \theta = \delta(p_i - p_j)$} \\ 
					& $p \in\mathbb{Z}$ &  \\
					\hline 
				\end{tabular}
				\caption{Scalar product to consider in a confined domain.}
				\label{tab:confined} 
			\end{table}

		\subsection{Boundary Conditions and Structure of the Solutions}
		
			As detailed in table~\ref{tab:confined} (left column), confining the wave field in a cylinder of radius $R$ and height $H$ reduces the accessible spatial domain, from $\mathbb{R}^+$ to $[0;\,R]$ in the radial direction, and from $\mathbb{R}$ to $[0;\,H]$ in the vertical direction. Such a change of geometry imposes a new set of constraints: contrary to infinite domains, the wave field now has to satisfy boundary conditions, namely zero orthogonal velocity on top and bottom at depth $H$
			\begin{equation}
				v_z(z=0) = 0 \mathrm{~~~~~~~and~~~~~~~} v_z(z=H) = 0, \label{sec5:eq1}
			\end{equation}
			as well as on the lateral cylindrical wall located at a radius $R$
			\begin{equation}
				v_r (r=R) = 0.\label{sec5:eq2}
			\end{equation}
			As opposed to the unbounded scenario detailed in the previous section, the full confinement induced by the lateral and horizontal boundary conditions leads to de-coupled vertical and horizontal dependence of the wave field as well as to a larger wave-wave interaction volume. The wave numbers and, consequently, the modes, allowed in such a confined geometry are now quantified: only a discrete collection of radial and vertical wave numbers can be selected.
			
			The first condition of equation~\eqref{sec5:eq1} is automatically fulfilled when writing the vertical dependence of the mode as a sine function with no phase shift, as previously assumed; the second condition can be solved analytically and, introducing $z^\star=mH$, it leads to
			\begin{equation}
				m H = z^\star = \frac{n \pi}{2}\mathrm{~~~~~~~with~~~~~~~} n \in \mathbb{N}.\label{sec5:eq3}
			\end{equation}
			As discussed in \cite{boury2018}, this vertical confinement and the condition stated by equation~\ref{sec5:eq3} produce a wave resonator through constructive and destructive interference, depending on the forcing wave frequency. While the forcing wave field might not fulfill this condition \textit{per se}, additional wave fields generated through non-linear interactions are compelled to satisfy it as soon as they fill the entire domain (see, e.g., generation of super-harmonics \citep{boury2020a}).

\begin{figure}
				\centering
				\includegraphics[scale=1]{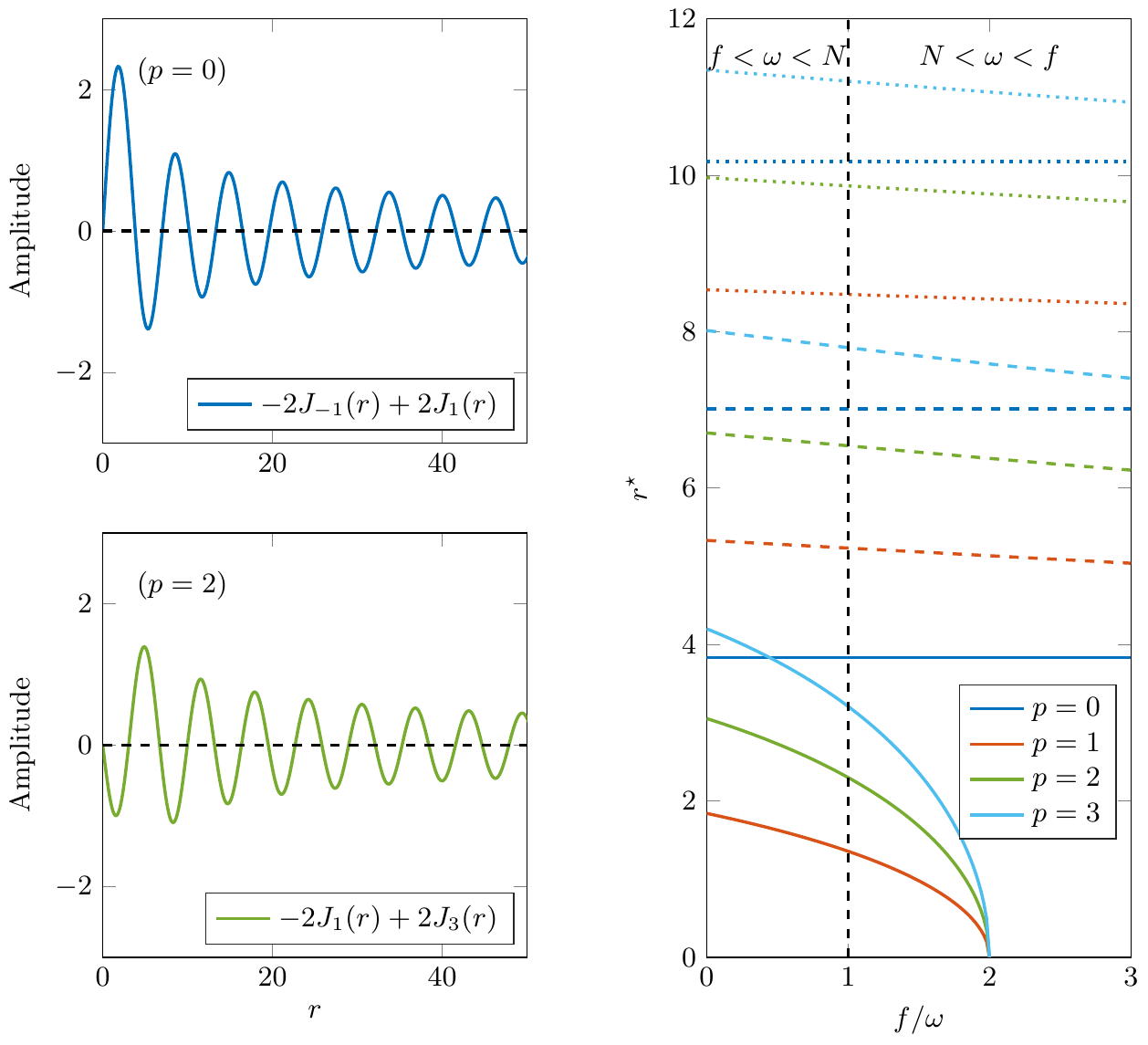}
				\caption{Illustration of condition~\eqref{sec5:eq4} imposed by the cylindrical boundary condition. Left: plots of~\eqref{sec5:eq4} for $f=0$ with $p=0$ (top) and $p=2$ (bottom). Right: locations $r^\star$ of the first (solid line), second (dashed line), and third (dotted line) non-zero nodes of~\eqref{sec5:eq5}, for $p$ from $1$ through $4$. The vertical dashed line helps distinguish the domains $f<\omega<N$ and $N<\omega<f$.}
				\label{fig:dispersionradiale}
			\end{figure}

			The cylindrical boundary is also constraining the allowed values of horizontal wave numbers through the non-penetration condition~\eqref{sec5:eq2} that can be written more explicitly
				\begin{equation}
					(f-2\omega) J_{p-1} (l R) + (f+2\omega) J_{p+1} (l R) = 0. \label{sec5:eq4}
				\end{equation}
				Contrary to the vertical condition~\eqref{sec5:eq3}, this equation shows that the horizontal description of the wave field, contained in the wave numbers $l$ and $p$, depends on the frequency for inertia-gravity waves. In the peculiar case of stratified non-rotating fluids, this relation no longer depends on $\omega$ and simply writes
				\begin{equation}
					J_{p-1} (l R) - J_{p+1} (l R) = 0. \label{sec5:eq5}
				\end{equation} 
				Note that, for axisymmetric modes ($p=0$), this condition reduces to
				\begin{equation}
					J_1 (l R) = 0, \label{sec5:eq6}
				\end{equation}
				for both gravity and inertial waves.	In a more general case, the zeros $l R = r^\star$ of equation~\eqref{sec5:eq4} can be determined numerically. We present, in figure~\ref{fig:dispersionradiale}, plots of the left-hand side of equation~\eqref{sec5:eq4} for $p=0$ (top left) and $p=2$ (bottom left), both for $f=0$, and the three first non-zeros solutions $r^\star$ as a function of $f/\omega$ (right). The colours stand for the value of $p$ from $0$ through $4$, and the solid, dashed, and dotted styles correspond to the first, second, and third solutions, respectively. A vertical dashed line at $f/\omega = 1$ indicates the cut-off between the gravity-dominated region ($f<\omega<N$) and the inertia-dominated region ($N<\omega<f$). As expected from the calculus performed with the axisymmetric assumption, for $p=0$ the solutions of equation~\eqref{sec5:eq5} do not depend on the frequency, but this is no longer the case as soon as $p\neq 0$.
	
			From now on, we should therefore consider box solutions, that we call modes, i.e. wave fields that comply both with the symmetry ($2\pi$-periodicity) and the geometry ($(r,z) \in [0;\,R] \times [0;\,H]$) of the system, leading to a discretisation of the allowed wave numbers. As shown before, defining the vertical velocity field is sufficent to describe these modes, and we write again
			\begin{equation}
				v_{z,j}(r,\theta,z,t) = v_{v,j}^0 J_{p_j} (l_j r) e^{i(\omega_j t - m_j z - p_j \theta)},
			\end{equation}
			where the values taken by $l_j$, $m_j$, and $p_j$, are now discrete. The scalar products on the radial and vertical coordinates defined in the previous section do not apply anymore for we have to take into account the finiteness of the domain and the discrete nature of the wave numbers. We present in table~\ref{tab:confined} (right column) the relevant scalar products that we will use to discuss the resonance conditions of these modes.
	
		\subsection{Resonance in Frequency}
		
			Similarly to the unconfined case, the temporal scalar product gives a resonance condition on the three wave frequencies that form a triad. In the absence of any additional constraint, this condition is always fulfilled and leads to the selection of two sub-harmonics of frequencies $\omega_1$ and $\omega_2$ such that $\omega_0 = \pm \omega_1 \pm \omega_2$. Note that this process is similar to the generation of super-harmonics in confined domains~\citep{boury2020a}.
	
		\subsection{Azimuthal Resonance}
		
			The scalar product on $\theta$ is the same as in the unconfined case, and therefore leads to the same resonance condition on the azimuthal wave numbers $p_0 = \pm p_1 \pm p_2$. Here, as in the unconfined case, the values of $p$ are integers, in order to comply with the $2\pi$-periodicity of the system.
			
		\subsection{Vertical Resonance}
		
			Due to the confinement, all of the vertical wave numbers $m$ can be expressed as ${m = n\pi / (2 H)}$ with $n\in\mathbb{N}$. Recasting the scalar product from $z \in \mathbb{R}$ with $m\in\mathbb{R}$ into $z\in [0;\,H]$ with $m$ defined through integers leads to the same resonance condition $m_0 = \pm m_1 \pm m_2$. Interestingly, although the vertical confinement of the wave field imposes a discrete set of vertical wave numbers, the vertical resonance condition can still be exact: this is due to the constant discrete spacing between two consecutive vertical wave numbers for modes in the confined domain, always distant of $\pi/(2 H)$ (see equation~\eqref{sec5:eq3}), that allows for vertical wave numbers that satisfy both the boundary conditions and the resonance relation.
	
		\subsection{Radial Resonance: Asymptotic Study}
		
			Along the radial direction, however, the difference is more significant. In order to discuss it we will use a similar asymptotic study as we did in the unconfined case. The scalar product performed on the linear part of the system of equations is now reduced to an integral from $0$ to $R$ (instead of $0$ to $+\infty$) where the wave numbers $l$ are discrete (instead of continuous). This imposes a rewriting of the integral over three Bessel functions, now radially limited in space, as
			\begin{equation}
				\forall (h,i,j)\in\mathbb{N}^3, \Xi_{hij} \simeq \frac{1}{R^2} \int_0^R \tilde{J}_h (l_0 r) \tilde{J}_i (l_1 r) \tilde{J}_j (l_2 r)  r \diff r,
			\end{equation}
			leading to a redefinition of the $\Gamma$ functions, given $(a,b,c,d)\in\mathbb{Z}^4$ and the radial interaction wave number $l_{bcd}$ defined in equation~\eqref{eq:lbcd}, as
			\begin{equation}
				\Gamma (a,b,c,d) = \frac{1}{R^2} \int_0^R \frac{1}{2\pi}\sqrt{\frac{l_{bcd}}{l_0 l_1 l_2}} \tilde{J}_{a+b+c+d} (l_{bcd} r ) \diff r.
			\end{equation}
			As for the unconfined case, thanks to some trigonometry, these functions can be written as a sum
			\begin{equation}
				\Gamma(a,b,c,d) = \Gamma^\mathsf{C} (a,b,c,d) + \Gamma^\mathsf{S} (a,b,c,d),
			\end{equation}
			with
			\begin{equation}
				\Gamma^\mathsf{C} (a,b,c,d) = \sqrt{\frac{2}{\pi^3 l_0 l_1 l_2}} \cos \left(\frac{\pi}{2}\left( a+b+c+d +\frac{1}{2}\right) \right) \frac{\mathsf{C}\left( l_{bcd}^\ast \right)}{l_{bcd}^\ast},
			\end{equation}
			and
			\begin{equation}
				\Gamma^\mathsf{S} (a,b,c,d) = \sqrt{\frac{2}{\pi^3 l_0 l_1 l_2}} \sin \left(\frac{\pi}{2}\left( a+b+c+d +\frac{1}{2}\right) \right) \frac{\mathsf{S}\left( l_{bcd}^\ast \right)}{l_{bcd}^\ast},
			\end{equation}
			where, for the sake of clarity, we use the notation
			\begin{equation}
				l_{bcd}^\ast = \sqrt{\frac{2 R | l_{bcd} |}{\pi}},
			\end{equation}
			for the reduced interaction radial wave number. Contrary to the unconfined case, in which the reduced sine and cosine Fresnel integrals are only evaluated in $0$ (if the resonance relation on $l$ is satisfied) or in $+\infty$ (if they are not), now they can be evaluated from $0$ to $l^\ast$, due to the finite size of the domain. Incidently, they are no longer equal to $0$ or to $1$, but to values that are continuously distributed in $[0;\,1]$. This allows for the radial resonance to be ``approximate'', i.e. $l_{bcd}\simeq 0$, without preventing the non-linear terms to be a second order forcing of the system.
			
		\subsection{Radial Resonance: Case Studies in Confined Domain}
			
			We now consider the same two case studies as in the unconfined geometry (see figure~\ref{fig:schematicradialexamples}), i.e. (1) a fully axisymmetric case $p_0 = p_1 = p_2 = 0$, and (2) a non-axisymmetric case $p_0 = 0$, $p_1=1$, and $p_2=-1$. The numerical investigation presented here will help discuss the impact of the finite size of the domain, or confinement of the wave fields, on the resonance.
			
			\subsubsection{Axisymmetric Case $p_0 = p_1 = p_2 = 0$}
			
\begin{figure}
				\centering
				\includegraphics[scale=1]{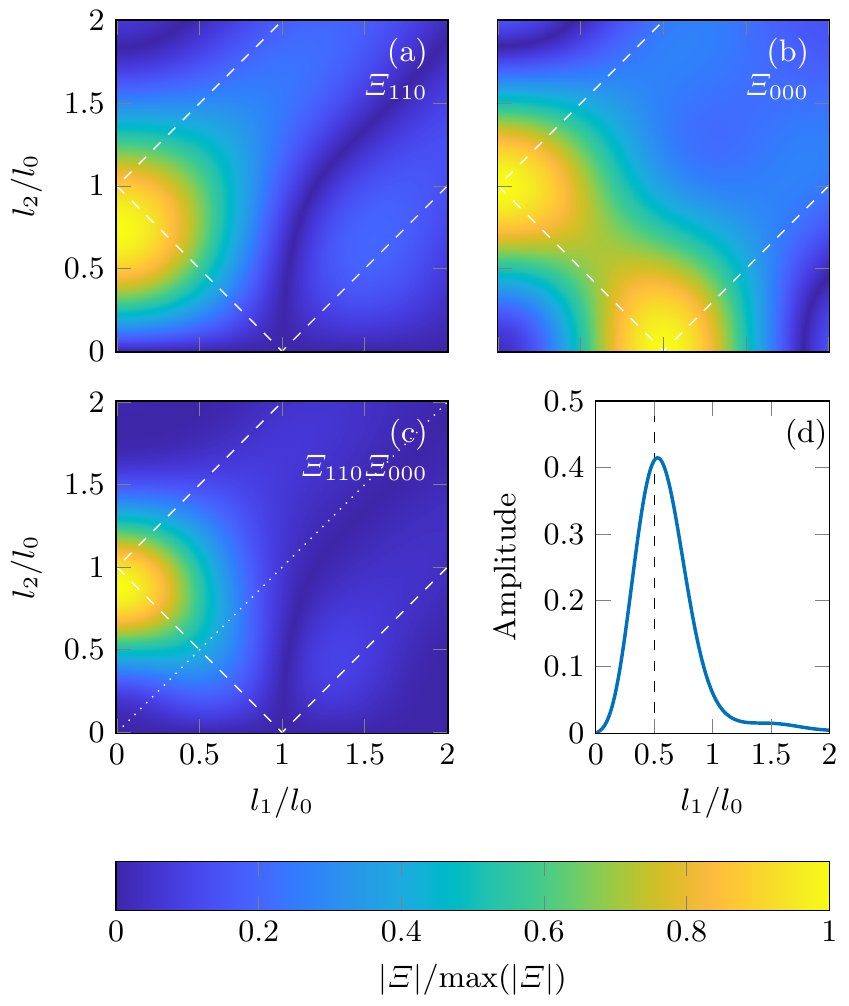}
				\caption{Colormaps of the normalised coefficients $|\Xi_{hij}|$ as a function of wave number ratios $l_1/l_0$ and $l_2/l_0$, with: (a) $|\Xi_{110}|$, (b) $|\Xi_{000}|$, and (c) product of the two previous quantities. Dashed lines show the locations of $l_0 \pm l_1 \pm l_2 = 0$. Plot (d) is the profile along the dotted line in plot (c).}
				\label{fig:cmaps3}
			\end{figure}

				As already discussed, in the fully axisymmetic case $p_0 = p_1 = p_2 = 0$, the radial scalar products of the non-linear term writes
				\begin{equation}
					\left\langle- (\mathbf{v} \cdot \mathbf{\nabla}) v_z\right\rangle_r = A_{z} \left[ M_{01} \Xi_{110} + M_{10} \Xi_{000} \right].
				\end{equation}
				The normalised absolute values of the corresponding coefficients $\Xi_{hij}$ are numerically computed and presented in figure~\ref{fig:cmaps3} as a function of $l_1/l_0$ and $l_2/l_0$, with $l_1/l_0$ and $l_2/l_0$ going from $0$ to $2$. We can see that, although $|\Xi_{110}|$ and $|\Xi_{000}|$ have different behaviours at a random location in the parameter space $(l_1/l_0,l_2/l_0)$, they generally show maximum values on the diagonals such that $\pm l_1 \pm l_2 = l_0$, shown by white dashed lines in figure~\ref{fig:cmaps1}. Their product is even more eloquent, as there is a clear maximum for $\pm l_1 \pm l_2 = l_0$ whereas the product is almost zero everywhere else. From these observations, we conjecture that the most likely values for radial wave numbers in TRI, for which relation~\eqref{eq4.16} has a non-zero right-hand side, satisfy the relation $l_0 = l_1+l_2$, as shown experimentally by~\cite{shmakova2018} and as already observed for Cartesian plane waves where it can be analytically demonstrated that $\pm l_1 \pm l_2 = l_0$ is a necessary condition \citep{joubaud2012}.
			
				For the sake of the demonstration, we shall clarify that having a non-zero product $|\Xi_{110}\Xi_{000}|$ is neither the only way for equation~\eqref{eq4.16} to be resonant (for example it is resonant if $\Xi_{000}$ is null and if $\Xi_{110}$ is not), nor does it ensures that this equation is resonant (depending on the values of $M_{01}$ and $M_{10}$, this equation can be non-resonant even if the product $|\Xi_{110}\Xi_{000}|$ is not null). Our reasoning nonetheless points towards a high probability of the system to select ``preferential'' configurations that correspond to the case of high value of $|\Xi_{110}\Xi_{000}|$, equivalent to large resonant terms and therefore large and efficient energy transfer. We note that, for the sub-harmonics to exist, the non-linear characteristic time (related to the growth of the instability) should overcome the viscous characteristic time (related to dissipative effects and therefore preventing the growth of sub-harmonics). This condition depends on the signs and values of the coefficients $M_{01}$ and $M_{10}$, that set the growth rate of the sub-harmonics, but the cases for which such condition is not satisfied are highly unlikely.
				
				As for comparison, similar colormaps to these presented in figure~\ref{fig:cmaps3} could be plotted for the non-linear terms in $2$D Cartesian geometry. In this case, the corresponding spatial integrals are associated to a product of three complex exponential functions (i.e. plane waves) instead of Bessel functions and, due to the properties of their scalar product, the colormaps are exactly $1$ over the diagonals $\pm l_1 \pm l_2 = l_0$ and $0$ everywhere else. Here, the finite size effect and approximate radial resonance (that can be seen through the more ``diffuse'' branches in the product in figure~\ref{fig:cmaps3}(c), compared to the unconfined case in figure~\ref{fig:cmaps1}), is due to the geometry of the wave field. 

			\subsubsection{Non-Axisymmetric Case $p_0 = 0$, $p_1=1$, and $p_2=-1$}
			
\begin{figure}
				\centering
				\includegraphics[scale=1]{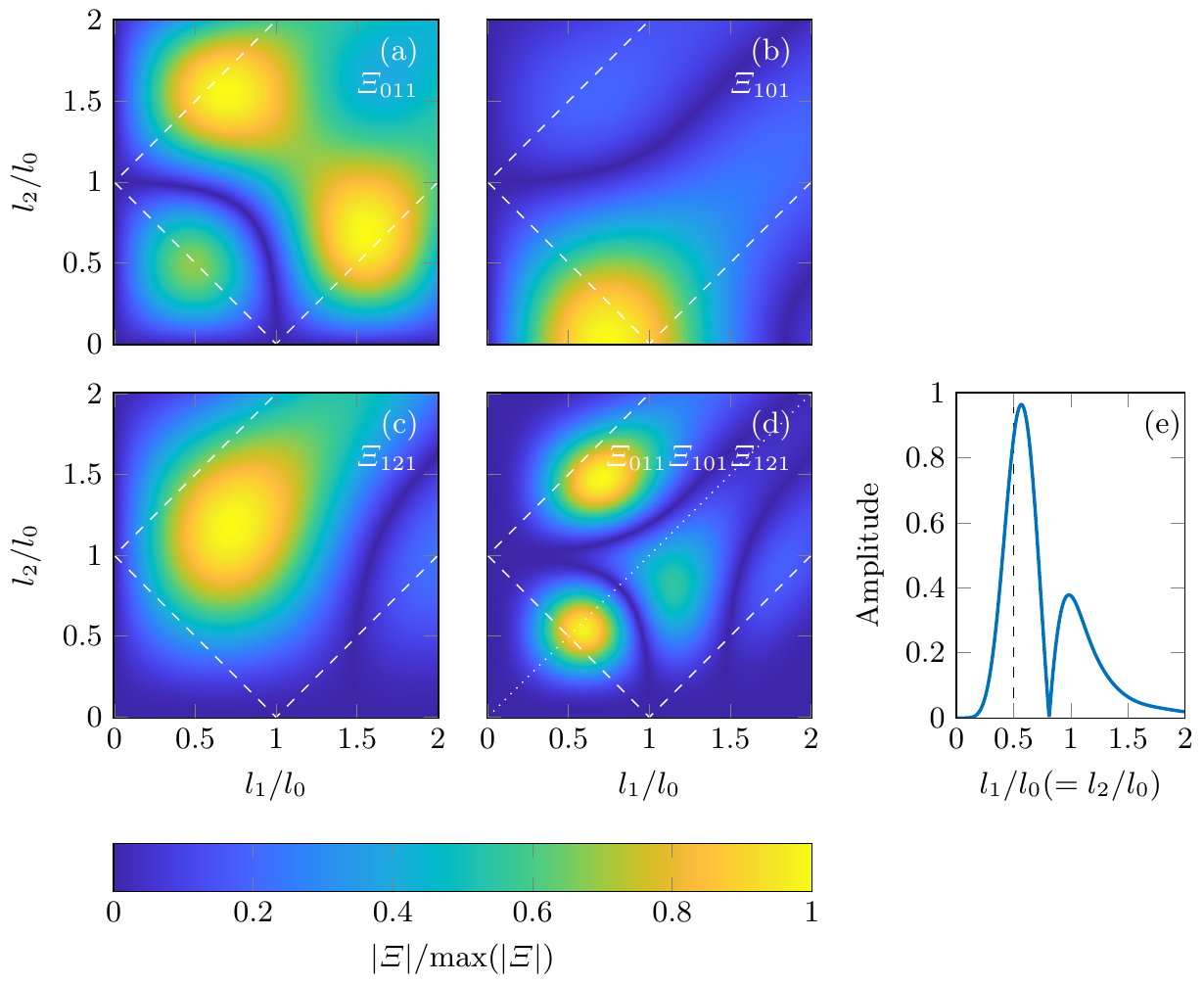}
				\caption{Colormaps of the normalised coefficients $|\Xi_{hij}|$ as a function of wave number ratios $l_1/l_0$ and $l_2/l_0$, with: (a) $|\Xi_{011}|$, (b) $|\Xi_{101}|$, (c) $|\Xi_{121}|$, and (d) product of the three previous quantities. Dashed lines show the locations of $l_0 \pm l_1 \pm l_2 = 0$. Plot (e) shows the profile along the dotted line in plot (d). }
				\label{fig:cmaps4}
			\end{figure}

				In the second case study with $p_0 = 0$, $p_1=1$, and $p_2=-1$, we have seen that the radial scalar product of the non-linear term writes
				\begin{equation}
					\left\langle- (\mathbf{v} \cdot \mathbf{\nabla}) v_z\right\rangle_r = A_{z} \left[ \frac{ M_{01}}{2}( \Xi_{101} - \Xi_{121}) - M_{10} \Xi_{011} \right].
				\end{equation}
				Figure~\ref{fig:cmaps4} presents colormaps of the normalised absolute value of the coefficients $\Xi_{hij}$, plotted as a function of $l_1/l_0$ and $l_2/l_0$, with $l_1/l_0$ and $l_2/l_0$ going from $0$ to $2$. These coefficients have, in general, maximum values on the diagonals given by $\pm l_1 \pm l_2 = l_0$ and there is a clear maximum for $\pm l_1 \pm l_2 = l_0$ for their product whereas it is almost zero everywhere else, leading to the same conclusion as in the fully axisymmetric case. We perform the same measure on the mid-height width of the branches to quantify the equality of the resonance relation. The profile presented in figure~\ref{fig:cmaps4}(e) is taken along the bissectrix shown by the dotted line in the product plot~\ref{fig:cmaps4}(d).
			
		\subsection{Approximated Triadic Resonance}
		
\begin{figure}
				\centering
				\includegraphics[scale=1]{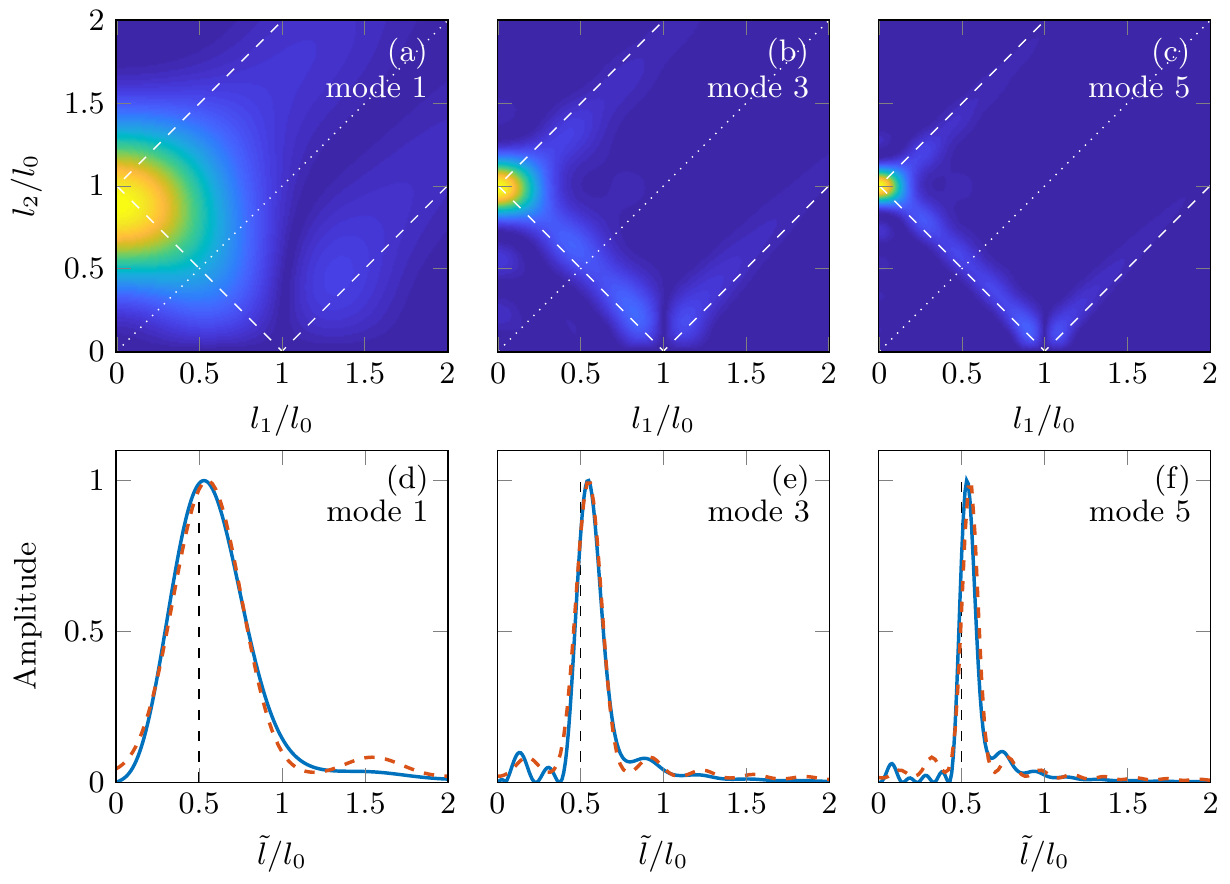}
				\caption{Top row: colormaps of the normalised coefficients $|\Xi_{110}\Xi_{000}|$ as a function of wave number ratios $l_1/l_0$ and $l_2/l_0$, with (a) $l_0=19\mathrm{~m^{-1}}$ (mode $1$), (b) $l_0=51\mathrm{~m^{-1}}$ (mode $3$), and (c) $l_0=82\mathrm{~m^{-1}}$ (mode $5$). Bottom row: measured profiles along the bissectrix corresponding to the white dashed line in the top row (blue solid curve) and prediction by the asymptotic theory (orange dashed line), for the same modes in (d), (e), and (f), respectively.}
				\label{fig:cmapsdeltal}
			\end{figure}

		We have seen, with the two case studies, that the radial resonance is not exact since the branches corresponding to exact resonant triads $l_0 \pm l_1 \pm l_2 = 0$ have a given spectral extension that we could quantify as a relative mid-height width $\Delta l / l_0$. This can be traduced as
			\begin{equation}
				l_0 \pm l_1 \pm l_2 = \varepsilon
			\end{equation}
			when the triad is radially resonant, with $\varepsilon \ll \Delta l$. The remaining question is to quantify this ``approximate'' resonance. In order to do so, we focus on the first bissectrix, i.e. the case $l_1=l_2$, and we introduce a common variable $\tilde{l} = l_1 = l_2$ to describe it. With this notation, we can define a renormalised variable $l^\star$ as
			\begin{equation}
				l^\star = \sqrt{\frac{2 l_0 R}{\pi} \left| 1 - \frac{2 \tilde{l}}{l_0}\right|},
			\end{equation}
			 According to our model, close to the radial resonance (i.e. for $l^\star$ close to $0$), the integrals $\Xi_{hij}$ are determined only by the reduced cosine Fresnel integral $\mathsf{C}(l^\star) / l^\star$ which should therefore fix the relative mid-height width $\Delta l /l_0$. To confirm the validity of our development, we present in figure~\ref{fig:cmapsdeltal} the colormaps of the product $\Xi_{110}\Xi_{000}$, corresponding to the first case study aforementioned, in three different cases: (a) $l_0=19\mathrm{~m^{-1}}$ (mode $1$), (b) $l_0=51\mathrm{~m^{-1}}$ (mode $3$), and (c) $l_0=82\mathrm{~m^{-1}}$ (mode $5$). We also present a comparison between the normalised profiles measured along the bissectrix (dotted line in plots (a) through (c)) and the cosine Fresnel integral $\mathsf{C}(l^\star)/l^\star$ predicted by the theory, for the three different values of $l_0$, in plots (c), (d), and (f), respectively. Since we are considering the product of two integrals $\Xi_{110}\Xi_{000}$, both behaving asymptotically as $\mathsf{C}(l^\star)/l^\star$, we plot the quadratic quantity $(\mathsf{C}(l^\star)/l^\star)^2$. Thanks to figures~\ref{fig:cmapsdeltal}(d,e,f), we note the very good agreement between the numerically computed profiles (blue solid lines) and the theory (orange dashed lines) to describe the behaviour close to the resonance located at $l_1/l_0 = l_2/l_0 = 0.5$. The mid-depth width, $\Delta l / l_0$, that can be extracted from the asymptotic theory is exactly the same as the one obtained from the numerics. As already discussed, these results can be extended to other configurations (e.g. second case study) and will lead to the same conclusions.
			
			Hence, our asymptotic theory predicts, in agreement to the exact computation of the resonant terms, that the radial resonance in such a cylindrical geometry is not exact and that we can quantitatively bound this approximativeness $\varepsilon$ of the resonance by a known $\Delta l$. For the cosine Fresnel integral, the relative mid-depth width is obtained when $l^\star=1$, which yields
			\begin{equation}
				\frac{\Delta l}{l_0} = \frac{\pi}{2 l_0 R}.
			\end{equation}
			Therefore, the relative mid-height width evolves as $(Rl_0)^{-1}$, i.e. the higher the mode the thiner the peak, exactly as observed in figures~\ref{fig:cmapsdeltal}(d,e,f). The relative mid-height width goes to zero as the order of the mode goes to infinity, corresponding to an exact resonance. By comparison, horizontal resonances for Cartesian plane waves are always exact; this result is recovered when considering high order radial modes in cylindrical geometry, when the area close to $r=0$ can be neglected and when the wave field can therefore be approximated by radially decreasing plane waves (such as $r \mapsto \cos(r) / r$).

		\subsection{Degrees of Freedom vs. Constraints}
			
			We proceed to a similar analysis as the one performed in unconfined domains. The sub-harmonics are, again, defined through $8$ parameters ($2\times 1$ frequencies and $2\times 3$ wave numbers). The constraints, however, are more numerous: $2$ dispersion relations, $4$ resonance conditions (TRI), and now $4$ additional constraints linked to boundary conditions ($2$ for each sub-harmonic). By analogy to the observations presented in \cite{boury2020a} for super-harmonics, we postulate that the constraints set by the boundary conditions prevail, and that the wave field always satisfies equations~\eqref{sec5:eq3} and~\eqref{sec5:eq4}, preferably to forming an exact triad. The reason for that is still the topic of ongoing research.  As a result, in addition to the internal wave dispersion relation, the frequencies and wave numbers are defined through
			\begin{eqnarray}
				\omega_0 &=& \pm \omega_1 \pm \omega_2,\\
				J_{p_1-1} (l_1 R) &=& J_{p_1 + 1} (l_1 R),\\
				J_{p_2-1} (l_2 R) &=& J_{p_2 + 1} (l_2 R),\\
				2 m_1 H &=& n_1 \pi, \\
				2 m_2 H &=& n_2 \pi.
			\end{eqnarray}

		\subsection{Experimental Observation}
		
\begin{figure}
			\centering
			\includegraphics[scale=1]{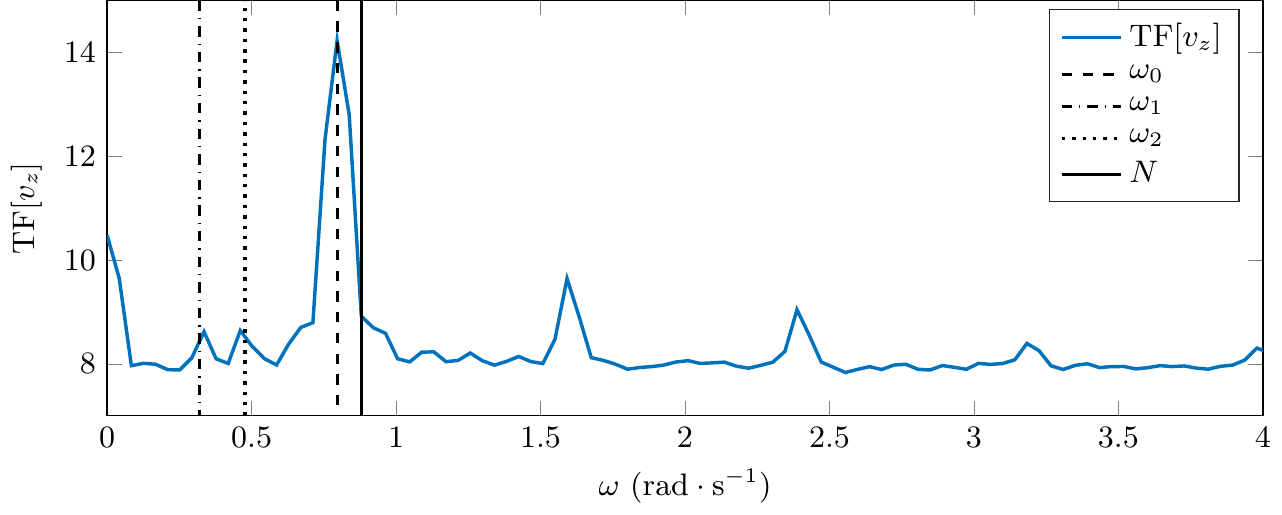}
			\caption{Fourier transform performed over the last two minutes of the experiment with a forcing at $\omega/N = 0.91$. From left to right: the dash-dotted line shows $\omega_1$, the dotted line shows $\omega_2$, both sub-harmonics, and the dashed line shows the forcing frequency $\omega_0$, and the solid line shows the buoyancy frequency $N$}
			\label{fig:confinedspectrum}
		\end{figure}

		We performed experiments for values of $\omega/N$ from $0.82$ to $0.92$, with a low amplitude ($a=2.5\mathrm{~mm}$) mode $1$ configuration at the generator~\citep{boury2018}. In several experiments, towards the end of the $10$ minute forcing one can observe the creation of sub-harmonics as presented in the spectrum in figure~\ref{fig:confinedspectrum} computed using the last two minutes of the acquisition. Two secondary waves are created at frequencies smaller than the imposed forcing ($\omega_1/N =0.36$ and $\omega_2/N=0.55$) that satisfy the triadic resonant condition $\omega_1 + \omega_2 = \omega_0$, as $\omega_1 = 0.4\,\omega_0$ and $\omega_2=0.6\,\omega_0$.
		
\begin{figure}
			\centering
			\includegraphics[scale=1]{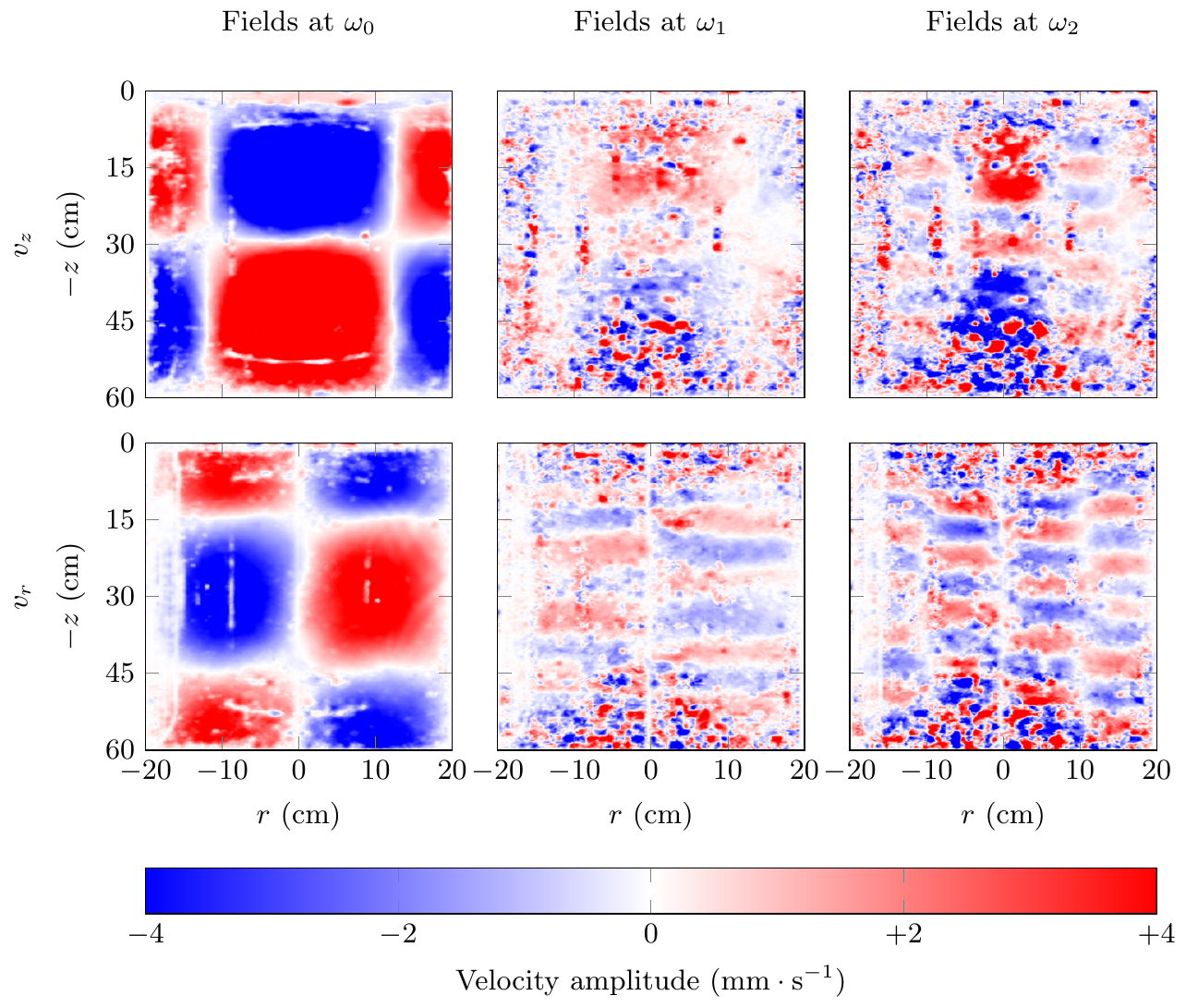}
			\caption{Vertical velocity (top row) and radial velocity (bottom row) after $9$ minutes of forcing in the experiment. From left to right, fileds are filtered at $\omega_0$, $\omega_1 = 0.4 \, \omega_0$, and $\omega_2 = 0.6\, \omega_0$.}
			\label{fig:confinedpiv}
		\end{figure}

		Filtered wave fields at the three frequencies $\omega_0$, $\omega_1$, and $\omega_2$, are presented in figure~\ref{fig:confinedpiv}, with the vertical velocity on top of the radial velocity. On the left, the primary wave shows high amplitude of about $4\mathrm{~mm\cdot s^{-1}}$, and is close to be a cavity mode $(1,2)$. The center and left columns are the two secondary waves identified from the spectrum in figure~\ref{fig:confinedspectrum}, at $\omega_1$ and $\omega_2$ respectively. We identify $1$ vertical wave length in the fields at $\omega_0$, $5$ in the fields at $\omega_1$, and $5$ or 6 in the fields at $\omega_2$: hence the resonance condition may not be satisfied for the vertical wave numbers, as we could have $m_0 \pm m_2 \pm m_1 \neq 0$. Nevertheless, this is not observed in all our experiments with sub-harmonic generation as sometimes we clearly have $m_0 = m_1 + m_2$, consistent with the vertical resonance relation~\eqref{sec5:eq3}. As regards the radial direction, we see different patterns in the filtered wave fields: the fields at $\omega_0$ and $\omega_1$ look like a radial mode $1$, but the field at $\omega_2$ looks like a radial mode $2$. This behaviour, however, is not a strong feature of the sub-harmonics generation \textit{via} TRI, as some experiments only show radial mode $1$ patterns.
		
		In general, we observe that in this confined configuration, the resonant conditions are not satisfied. As explained previously, the reason is that the selected frequencies and wave lengths are constrained by the boundary conditions. This is supported by our experimental observations, as can be seen in figure~\ref{fig:confinedpiv}, in the vertical plane, the sub-harmonics can be identified as cavity modes. Using the cavity mode formalism detailed in \cite{boury2020a} we see that, for example, in figure~\ref{fig:confinedpiv}, the field at $\omega_1$ is a mode $(1,10)$ and the field at $\omega_2$ is a mode $(2,10)$. Results already derived by \cite{boury2020a} on super-harmonic generation can be extended to this problem: the predicted frequencies associated to these cavity modes $(1,10)$ and $(2,10)$ are $\omega_{1,10}/N=0.34$ and $\omega_{2,12}/N=0.56$, close to the experimental values of $\omega_1/N=0.36$ and $\omega_2/N=0.55$.

		Contrary to the observations of \cite{maurerPhD} in his focusing experiments, but in agreement with \cite{shmakova2018}, we did not see any axisymmetry breaking in our experimental wave fields. Due to the poor visualisation in the horizontal plane, however, this statement could not be further explored. Our conjecture is that the presence of a cylindrical boundary at fixed radius might prevent the secondary waves from breaking the symmetry, contrary to Maurer's observations in wave focusing experiments in which non-zero radial velocity was detected~\citep{maurerPhD}: such velocities, indeed, could not be described by the Bessel functions we used and could be contradictory to the condition of zero radial velocity at the cylindrical bound.

		The primary and secondary waves are not always satisfying the TRI relations on the wave numbers, particularly the radial wave number, which means that the boundary conditions that set the cavity mode are, in that sense, ``stronger'' than the resonance conditions. In addition, part of the velocity fields are blurred, for example the vertical velocity at $\omega_1$ and $\omega_2$ close to $r=0\mathrm{~cm}$ at the top and at the bottom of the tank (figure~\ref{fig:confinedpiv}). This is likely due to the presence of other modes or to exchanges between the cavity mode and another wave field set by the TRI conditions. Similarly to Super-Harmonic Generation, the existence of resonance conditions in TRI may prescribe the cavity modes allowed for non-linear interaction.

	\section{Conclusions and Discussion}
	\label{sec:conclusions}
		
			In the present study, we investigated the conditions upon which resonant triads can exist in three dimensional axisymmetric geometry. In addition to the analytical and numerical investigation of the resonance conditions and of the coercion by boundary conditions, we provided experimental illustrations of the spontaneous generation of sub-harmonics in confined and unconfined domains in order to validate the theory. Two specific cases have been studied, defined according to the physical domain accessible to the waves: the unconfined case and the confined case (see table~\ref{tab:domains}).
		
			We have demonstrated that, in both cases, the sub-harmonic frequencies satisfy a resonance condition
			\begin{equation}
				\omega_0 = \pm \omega_1 \pm \omega_2,
			\end{equation}
			therefore forming a triad. The investigation of the spatial structure, however, yields different results depending on the confinement of the wave field. In the unconfined case, the three dimensional spatial structure is prescribed by resonance conditions on the wave numbers, as follows
			\begin{eqnarray}
				l_0 &=& \pm l_1 \pm l_2,\\
				p_0 &=& \pm p_1 \pm p_2,\\
				m_0 &=& \pm m_1 \pm m_2.
			\end{eqnarray}
			Conversely, in the confined case, the spatial structure is primarily constrained by boundary conditions, meaning that the wave numbers satisfy
			\begin{eqnarray}
				J_{p_1-1} (l_1 R) &=& J_{p_1 + 1} (l_1 R),\\
				J_{p_2-1} (l_2 R) &=& J_{p_2 + 1} (l_2 R),\\
				2 m_1 H &=& n_1 \pi, \\
				2 m_2 H &=& n_2 \pi,
			\end{eqnarray}
			and that the triad is not necessarily spatially resonant -- as discussed with theoretical derivations and confirmed by experimental observations. The existence of sub-harmonics that do not fully satisfy the resonance conditions may be related to quasi-resonances and viscous effects that allow for a larger resonance bandwidth (see, e.g., the results on three-wave interactions for capillary waves of \cite{cazaubiel2019}). A similar discrimination between confined and unconfined domains is expected for $2$D and $3$D spatially resonant triads or cavity modes in Cartesian geometry. Nonetheless, the reason why boundary conditions prevail over triadic equations, and the exact process of modal selection within the cavity, remain unknown. A possible explanation is that the triadic interaction remains a local process occuring at the most energetic locations (as discussed in \cite{boury2020b} in the case of unstable branches of a wave attractor) but, as the sub-harmonic wave field develops and progressively fills the confined domain, it adjusts to the global boundary conditions.
			
			The singular difference between the confined and unconfined domain lies, therefore, in the radial and vertical spatial directions, as stated in table~\ref{tab:domains}, since introducing boundary conditions on these two directions have very different implications. On the one hand, the discretization induced by the vertical confinement does not prevent the vertical wave numbers to be resonant: we note that, if there exists an integer $n_0$ such that $2 m_0 H = n_0 \pi$, we can always find two integers $n_1$ and $n_2$ such that the resonance condition $m_0 = \pm m_1 \pm m_2$ is fulfilled with $2 m_1 H = n_1 \pi$ and $2 m_2 H = n_2 \pi$, so that all vertical wave numbers satisfy the boundary conditions. On the other hand, the discrete selection of radial wave numbers is usually incompatible with an exact radial resonance. Mathematically, this issue arises while integrating the $\Xi_{hij}$ functions previously defined, since one of the bound is defined to be $R \times l_{bcd}$ -- where $R$ is the size of the domain and $l_{bcd}$ is defined as the linear combination of radial wave numbers. In the unconfined case, this quantity $R \times l_{bcd}$ is exactly equal to $0$ if and only if $l_{bcd}=0$ (i.e. an exact resonance is reached) and equal to $+\infty$ otherwise, since the size of the domain $R$ is infinite; this leads to integrated quantities (the Fresnel functions, which are a proxy to quantify the non-linear interaction between the waves) that are exactly evaluated in $0$ for resonant triads and in $+\infty$ otherwise. In confined domains, the size of the domain being finite, the quantity $R \times l_{bcd}$ is close to $0$ if a quasi resonance $l_{bcd}\simeq 0$ is reached, and takes high values when the waves of the triad are far from being resonant; this allows for quasi-resonances, as the integrated quantities (once again, the Fresnel integrals), are evaluated close to $0$ for such quasi-resonance triads -- a situation not allowed in the case of an unconfined domain. We note that, given this discussion, the bandwidth allowing for approximate resonances goes as $1/R$, meaning that at fixed radial wave numbers, the larger the domain, the more likely are the resonances to be exact.
			
			An important finding is the possibility, while generating sub-harmonics though TRI, of symmetry breaking: the secondary waves can have non-zero orthoradial wave numbers although the primary wave is axisymmetric, leading to the generation of counter-rotating wave fields. These wave numbers still have to satisfy a resonance condition. Our experiment shows $2\pi$-periodic cylindrical sub-harmonics, but higher periodicities could exist in such a non-linear process.
			
			Due to mathematical complexities, the cylindrical radial resonance condition is not analytically demonstrated over the whole domain, but only verified asymptotically, and we observe that it is consistent with experimental data. A rigorous proof of the exact equality of this resonance condition is still a challenge for future research. In addition, we should point out that the explanation of symmetry breakings such as identified by~\cite{maurerPhD} in the case of high amplitude non-linear interactions could lie in the calculus of the growth rates of the triadic instability, which is beyond the scope of the present study.
			
			\begin{table}
			\centering
			\begin{tabular}{ccc}
			\hline 
			~ & Unconfined case & Confined case \\ 
			\hline 
			Temporal ($t$) & $-\infty$ to $+\infty$ & $-\infty$ to $+\infty$ \\ 
			\hline 
			Radial ($r$) & $0$ to $+\infty$ & $0$ to $R$ \\ 
			\hline 
			Vertical ($z$) & $-\infty$ to $+\infty$ & $0$ to $L$ \\ 
			\hline 
			Azimuthal ($\theta$) & $0$ to $2\pi$ & $0$ to $2\pi$ \\ 
			\hline 
			\end{tabular}
			\caption{Comparison of the physical domains between the unconfined case and the confined case. \label{tab:domains}}  
			\end{table}

	\appendix
	
	\section{Non-Linear Terms}
	\label{app:nonlinearterms}
	
		We present here the computation of the non-linear advection terms. First of all, we perform a direct calculation of the advection term along $z$, as follows
		\begin{eqnarray}
			- \left( \mathbf{v} \cdot \mathbf{\nabla} \right) v_{z} &=& - \sum_{i=1}^3 \sum_{j=1}^3 \left( \mathbf{v}_i \cdot \mathbf{\nabla} \right) v_{z,j} \\
			&=& -\sum_{i=1}^3 \sum_{j=1}^3 \left[ v_{r,i}\partial_r v_{z,j} + \frac{v_{\theta,i}}{r}\partial_\theta v_{z,j} + v_{z,i}\partial_z v_{z,j} \right] \\
			&=& -\sum_{i=1}^3 \sum_{j=1}^3 i v_{z,i}^0 v_{z,j}^0 \left[ -\frac{m_j l_i}{ l_j} J_{p_j}' J_{p_i}' + \frac{m_j p_j p_i}{ l_j^2 r^2} J_{p_j} J_{p_i} - m_i J_{p_i}J_{p_j}\right] \Phi_{ij}\\
			&=& -\sum_{i=1}^3 \sum_{j=1}^3 i v_{z,i}^0 v_{z,j}^0 \left[ -\frac{m_j l_i}{4 l_j} \left(J_{p_j-1} - J_{p_j+1} \right) \left(J_{p_i-1} - J_{p_i+1}\right)\right. \\ 
			&~&~+ \left.\frac{m_j l_i}{4 l_j} \left(J_{p_j-1} + J_{p_j+1} \right) \left(J_{p_i-1} + J_{p_i+1}\right) - m_i J_{p_i}J_{p_j}\right] \Phi_{ij}\\
			&=& -\sum_{i=1}^3 \sum_{j=1}^3 i \frac{v_{z,i}^0 v_{z,j}^0}{l_j} \left[ \frac{M_{ij}}{2} \left( \mathsf{J}_{p_i -1}^{p_j+1} + \mathsf{J}_{p_i +1}^{p_j-1} \right) - M_{ji} \mathsf{J}_{p_i}^{p_j}\right] \Phi_{ij}
		\end{eqnarray}
	
		We now compute the advection terms for the other components of the velocity field, and for the buoyancy. The radial velocity term needs to be integrated to recover the projection along $z$ of the advection term, with an additional $\epsilon_{i,j}$ term that we will neglect
		\begin{eqnarray}\hspace*{-1cm}
			-\int \left( \mathbf{v} \cdot \mathbf{\nabla} \right) v_{r} &=& -\sum_{i=1}^3 \sum_{j=1}^3 \int \left[ v_{r,i} \partial_r v_{r,j} + \frac{v_{\theta,i}}{r} \partial_\theta v_{r,j} + v_{z,i} \partial_z v_{r,j} - \frac{v_{\theta,i} v_{\theta,j}}{r}\right] \diff r\\
			&=& -\sum_{i=1}^3 \sum_{j=1}^3 -i \frac{m_j}{l_j^2} \int \left[v_{r,i} \partial^2_r v_{z,j} + v_{z,i} \partial_z \partial_r v_{z,j} + \frac{v_{\theta,i}}{r} \partial_r \partial_\theta v_{z,j} - \frac{v_{\theta,i} \partial_\theta v_{z,j}}{r^2} \right]\diff r\\
			&=& -\sum_{i=1}^3 \sum_{j=1}^3 -i \frac{m_j}{l_j^2} \left[ v_{r,i} \partial_r v_{z,j} + v_{z,i} \partial_z v_{z,j} + \frac{v_{\theta,i}}{r} \partial_\theta v_{z,j} \right] \\
			&~&~ + i \frac{m_j}{l_j^2} \int \left[  \partial_r v_{r,i} \partial_r v_{z,j} + \partial_r v_{z,i} \partial_z v_{z,j} + \partial_r \left(\frac{v_{\theta,i}}{r}\right)  \partial_\theta v_{z,j} + \frac{v_{\theta,i} \partial_\theta v_{z,j}}{r^2} \right]\diff r\\
			&=& -\sum_{i=1}^3 \sum_{j=1}^3 -i \frac{m_j}{l_j^2} \left( \mathbf{v}_i \cdot \mathbf{\nabla} \right) v_{z,j} + i \frac{m_j}{l_j^2} \int \left[  \partial_r v_{r,i} \partial_r v_{z,j} + \partial_r v_{z,i} \partial_z v_{z,j} + \frac{\partial_r v_{\theta,i}}{r}  \partial_\theta v_{z,j}  \right]\diff r\\
			&=& -\sum_{i=1}^3 \sum_{j=1}^3 -i \frac{m_j}{l_j^2} \left( \mathbf{v}_i \cdot \mathbf{\nabla} \right) v_{z,j} - \int \left[  \partial_r v_{r,i} v_{r,j} + \partial_r v_{\theta,i}  v_{\theta,j} - i \frac{m_j}{l_j^2}  \partial_r v_{z,i} \partial_z v_{z,j} \right]\diff r\\
			&=& -\sum_{i=1}^3 \sum_{j=1}^3 -i \frac{m_j}{l_j^2} \left( \mathbf{v}_i \cdot \mathbf{\nabla} \right) v_{z,j} -  \frac{1}{2} \left( v_{r,i} v_{r,j} + v_{\theta,i} v_{\theta,j} \right) + i \frac{m_j}{l_j^2} \int \partial_r v_{z,i} \partial_z v_{z,j} \diff r\\
			&=& -\sum_{i=1}^3 \sum_{j=1}^3 -i \frac{m_j}{l_j^2} \left( \mathbf{v}_i \cdot \mathbf{\nabla} \right) v_{z,j} + \epsilon_{i,j}
		\end{eqnarray}
		
		The azimuthal velocity term has to be multiplied by $r$ to cancel out the dependence in $1/r$, giving
		\begin{eqnarray}
			- r \left( \mathbf{v} \cdot \mathbf{\nabla} \right) v_{\theta} &=& - \sum_{i=1}^3 \sum_{j=1}^3 r \left[ v_{r,i} \partial_r v_{\theta,j} + \frac{v_{\theta,i}}{r} \partial_\theta v_{\theta,j} + v_{j,i} \partial_z v_{\theta,j} + \frac{v_{r,i}v_{\theta,j}}{r} \right] \\
			&=& - \sum_{i=1}^3 \sum_{j=1}^3 \left[\left( r v_{r,i} \partial_r v_{\theta,j} + v_{r,i} v_{\theta,j} \right) + v_{\theta,i}\partial_\theta v_{\theta,j} + r v_{z,i} \partial_z v_{\theta,j} \right] \\
			&=& - \sum_{i=1}^3 \sum_{j=1}^3 \left[ v_{r,i} \partial_r \left( r v_{\theta,j} \right) + \frac{v_{\theta,i}}{r} \partial_\theta \left( r v_{\theta,j} \right) + v_{z,i} \partial_z \left( r v_{\theta,j} \right) \right] \\
			&=& - \sum_{i=1}^3 \sum_{j=1}^3 \frac{m_j p_j}{l_j^2} \left[ v_{r,i} \partial_r v_{z,j} + \frac{v_{\theta,i}}{r} \partial_\theta v_{z,j} + v_{z,i} \partial_z v_{z,j} \right] \\
			&=& -  \sum_{i=1}^3 \sum_{j=1}^3 \frac{m_j p_j}{l_j^2} \left( \mathbf{v}_i \cdot \mathbf{\nabla} \right) v_{z,j}
		\end{eqnarray}
		
		Finally, the buoyancy term is directly proportional to the vertical velocity, so that
		\begin{eqnarray}
			- \left( \mathbf{v} \cdot \mathbf{\nabla} \right) b &=& - \sum_{i=1}^3 \sum_{j=1}^3 \left( \mathbf{v}_i \cdot \mathbf{\nabla} \right) \left( i \frac{N^2}{\omega_j} v_{z,j} \right) = -  \sum_{i=1}^3 \sum_{j=1}^3  i \frac{N^2}{\omega_j} \left( \mathbf{v}_i \cdot \mathbf{\nabla} \right) v_{z,j}
		\end{eqnarray}

	\section{Fresnel Integrals}
	\label{sec:app:fresnel}
	
\begin{figure}
			\centering
			\includegraphics[scale=1]{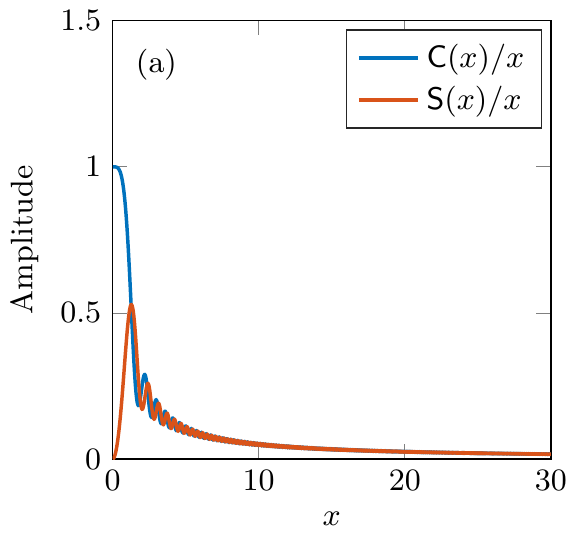}
			\includegraphics[scale=1]{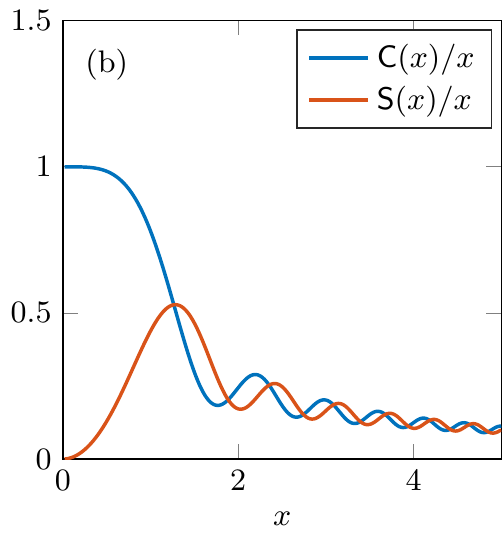}
			\caption{Plots of the modified cosine and sine Fresnel integrals, for (a) $x \in [0;~30]$ and (b) $x \in [0;~5]$.}
			\label{fig:app2}
		\end{figure}

		For the sake of the discussion, we recall the definition of the cosine and sine Fresnel integrals, $x \mapsto \mathsf{C}(x)$ and $x \mapsto \mathsf{S}(x)$, respectively, for $x \in \mathbb{R}$ \citep{NIST2010}
		\begin{equation}
			\mathsf{C}(x) = \int_0^x \cos \left( \frac{1}{2} \pi t^2 \right) \diff t \mathrm{~~~~~~~and~~~~~~~} \mathsf{S}(x) = \int_0^x \sin \left( \frac{1}{2} \pi t^2 \right) \diff t.
		\end{equation}
		The ``modified'' cosine and sine Fresnel integrals, defined by $x \mapsto \mathsf{C}(x) / x$ and $x \mapsto \mathsf{S}(x) / x$, respectively, for $x \in \mathbb{R}$, and used in the asymptotic developments aforementioned, are plotted in figure~\ref{fig:app2}(a) for $x \in [0;~30]$ with a zoomed-in version in figure~\ref{fig:app2}(b) for $x \in [0;~5]$. The values at $x=0$ ($0$ and $1$ for the modified cosine and sine Fresnel integrals, respectively) can be clearly seen, as well as the rapid decay towards zero for larger values of $x$.
			
\bigskip
\textbf{Acknowledgments}
	
		This work has been funded through grant ANR-17-CE30-0003 (DisET). Data processing has been made possible thanks to the ressources of the PSMN based at the ENS de Lyon. The authors gratefully acknowledge P.~Meunier and L.R.R.~Maas for fruitful discussions and inputs. S.B. wants to thank the labex iMust and the IDEX Lyon for supporting his research and travels.
	

%
\bibliographystyle{jfm}
\bibliography{bibliotriadic}

\end{document}